\documentclass[useAMS,usenatbib]{mn2e}

\usepackage[latin1]{inputenc}
\usepackage{graphicx}
\usepackage{epstopdf}

\usepackage{rotate}
\usepackage{amsmath, float}
\usepackage{amssymb}
\usepackage{soul}
\usepackage{txfonts}
\usepackage{fleqn}
\usepackage{color}
\usepackage{comment}

\usepackage{mathrsfs}  
\usepackage{booktabs}
\usepackage{colortbl} 

\newcolumntype{G}{>{\columncolor[gray]{0.8}}l} 

\newcommand{\be}{\begin{equation}}
\newcommand{\ee}{\end{equation}}
\newcommand{\bdm}{\begin{displaymath}}
\newcommand{\edm}{\end{displaymath}}
\newcommand{\bea}{\begin{multline}}
\newcommand{\eea}{\end{multline}}
\newcommand{\ba}{\begin{align}}
\newcommand{\ea}{\end{align}}

\newcommand{\thi}{\theta_{\rm i} }

\def\simlt{\mathrel{\hbox{\rlap{\hbox{\lower4pt\hbox{$\sim$}}}\hbox{$<$}}}}
\def\simgt{\mathrel{\hbox{\rlap{\hbox{\lower4pt\hbox{$\sim$}}}\hbox{$>$}}}}

\title[Magnetic Fields in Bow-Shock Pulsar Wind Nebulae]{A Laminar
  Model for the Magnetic Field Structure in Bow-Shock
Pulsar Wind Nebulae}
\author[N. Bucciantini]{
N. Bucciantini$^{1,2,3}$\thanks{E-mail: niccolo@arcetri.astro.it} \\
$^{1}$INAF - Osservatorio Astrofisico di Arcetri, Largo E. Fermi 5,
I-50125 Firenze, Italy\\
$^{2}$Dipartimento di Fisica e Astronomia, Universit\`a degli Studi di Firenze, Via G. Sansone 1, 
I-50019 Sesto F.~no  (Firenze), Italy\\
$^{3}$INFN - Sezione di Firenze, Via G. Sansone 1, I-50019 Sesto F.~no  (Firenze), Italy}

\begin{document}
 
\date{Accepted / Received}

\maketitle

\label{firstpage}

\begin{abstract}
Bow Shock Pulsar Wind Nebulae are a class of non-thermal sources, that
form when the wind of a pulsar moving at supersonic speed interacts
with the ambient medium, either the ISM or in a few cases the cold ejecta of the
parent supernova. These systems have attracted attention in recent
years, because they allow us to investigate the properties of the
pulsar wind in a different environment from that of canonical Pulsar Wind
Nebulae in Supernova Remnants. However, due to the complexity of the
interaction, a full-fledged multidimensional analysis is still
laking. We present here a simplified approach, based on Lagrangian tracers, to model the magnetic
field structure in these systems, and use it to compute the magnetic
field geometry, for various configurations in terms of relative
orientation of the magnetic axis, pulsar speed and observer
direction. Based on our solutions we have computed a set of radio emission
maps, including polarization, to investigate the variety of possible
appearances, and how the observed emission pattern can be used to
constrain the orientation of the system, and the possible
presence of turbulence.
\end{abstract}

\begin{keywords}
 MHD - radiation mechanisms: non-thermal - polarization - relativistic
 processes- ISM: supernova remnants 
\end{keywords}

\section{Introduction}

Pulsars are rapidly rotating and strongly magnetized neutron stars \citep{Gold68a,Pacini67a}. Due
to the interplay of a fast rotation and a strong magnetic field, a
large induced
electric field forms at their surface, capable of pulling electrons
out of the crust \citep{Goldreich_Julian69a}, and driving a pair creation cascade process, that
leads to the formation of a pair dominated outflow, known as {\it
  pulsar wind} \citep{Ruderman_Sutherland75a,Arons_Scharlemann79a,Hibschman_Arons01b,Timokhin_Arons13a}. This wind is accelerated beyond the Light Cylinder to
high Lorentz factors. While the mechanism responsible of such
acceleration is still poorly understood \citep{Michel73b,Bogovalov99a,Kirk_Skjaeraasen03a,Mochol17a}, there is no doubt that at
the typical distances where this wind is seen to interact with the
environment, it is cold and likely endowed with a toroidal magnetic
field, wound up around the pulsar spin-axis, and carrying a non-negligible
fraction of the pulsar spin-down luminosity \citep{Melatos98a}.

The interaction of the pulsar wind with the surrounding environment
gives rise to a {\it Pulsar Wind Nebula} (PWN) \citep{Gaensler_Slane06a,Bucciantini08b,Olmi_Del-Zanna+16a}. A bubble of relativistic
pairs and magnetic field, that shines with a non-thermal broad band
spectrum, extending from radio to X-rays and $\gamma$-rays, via
synchrotron and inverse Compton emission. The Crab nebula
is the prototype of PWNe inside the parent Supernova Remnant (SNR) \citep{Hester08a}. Given
that the typical pulsar speed at birth, the so called {\it kick
  velocity}, is usually of the order of $100-400$ km s$^{-1}$ \citep{Cordes_Chernoff98a,Arzoumanian_Chernoff+02a,Sartore_Ripamonti+10a,Verbunt_Igoshev+17a}, much
smaller than the expansion speed of young SNRs [typically a few
thousands  km s$^{-1}$ \citep{Truelove_McKee99a,Hughes99a,Hughes00a,DeLaney_Rudnick03a,Borkowski_Reynolds+13a,Tsuji_Uchiyama16a}], for the first few thousands of years, the PWN
is going to remain confined within the parent SNR \citep{van-der-Swaluw_Achterberg+03a,van-der-Swaluw_Downes+04a,Temim_Slane+15a}. However the SNR
expansion speed will drop in time as the expansion proceeds \citep{Cioffi_McKee+88a,Leahy_Green+14a,Sanchez-Cruces_Rosado+18a}, sweeping
up more and more ISM material, such that on a typical timescale of a
few tens of thousands of years, the pulsar can escape from the SNR,
and begin to interact directly with the ISM.  Given the
typical sound speed of the ISM, the pulsar motion is highly supersonic,
and the ram pressure balance between the pulsar wind and the ISM flow
gives rise to a cometary-like nebula \citep{Wilkin96a} known as Bow-Shock Pulsar Wind
Nebula (BSPWN), of shocked pulsar and ISM material \citep{Bucciantini_Bandiera01a,Bucciantini02b}. If the ISM is cold and its neutral fraction is high,
these nebulae can be observed in H$_\alpha$
emission \citep{Kulkarni_Hester88a,Cordes_Romani+93a,Bell_Bailes+95a,van-Kerkwijk_Kulkarni01a,Jones_Stappers+02a,Brownsberger_Romani14a,Romani_Slane+17a}, due to charge exchange and collisional
excitation, taking place in the shocked ISM downstream of the forward
bowshock
\citep{Chevalier_Kirshner+80a,Hester_Raymond+94a,Bucciantini_Bandiera01a,Ghavamian_Raymond+01a},
or alternatively  in the UV \citep{Rangelov_Pavlov+16a}, and IR \citep{Wang_Kaplan+13a}. On the other hand the shocked pulsar material is
expected to emit non-thermal synchrotron radiation, and indeed many
such systems have been identified in recent years either in radio or
in X-rays \citep{Arzoumanian_Cordes+04a,Kargaltsev_Pavlov+17a,Kargaltsev_Misanovic+08a,Gaensler_van-der-Swaluw+04a,Yusef-Zadeh_Gaensler05a,Li_Lu+05a,Gaensler05a,Chatterjee_Gaensler+05a,Ng_Camilo+09a,Hales_Gaensler+09a,Ng_Gaensler+10a,De-Luca_Marelli+11a,Marelli_De-Luca+13a,Jakobsen_Tomsick+14a,Misanovic_Pavlov+08a,Posselt_Pavlov+17a,Klingler_Rangelov+16a,Ng_Bucciantini+12a}. There is also an interesting dichotomy between
H$_\alpha$ emitting systems and synchrotron emitting ones, suggesting
that energetic pulsars, more likely to drive bright synchrotron
nebulae, might pre-ionize the incoming ISM suppressing line emission from neutral
hydrogen. 

Despite the fact that the study of BSPWNe dates back to almost two
decades ago \citep{Cordes_Romani+93a,Bucciantini_Bandiera01a,Bucciantini_Amato+05a}, in recent years there has been a renewed
interest in these systems. In part due to an increasing number of
discoveries in radio and X-rays, that have enlarged our sample and
shown us the large variety of morphologies characterizing them, in part because pulsars are perhaps
the most efficient antimatter factories in the Galaxy, and BSPWNe have
been suggested as one of the major contributors to the positron excess
observed by PAMELA
\citep{Adriani_Barbarino+09a,Hooper_Blasi+09a,Blasi_Amato11a,Adriani_Bazilevskaya+13a,Aguilar_Alberti+13a},
in competition with dark matter  \citep{Wang_Pun+06a}. In recent years the discovery of peculiar X-ray outflows
in BSPWNe \citep{Hui_Huang+12a,Pavan_Bordas+14a} has also put into question or current
understanding of pulsar winds \citep{Spitkovsky06a,Tchekhovskoy_Spitkovsky+13a,Tchekhovskoy_Philippov+16a}, and the canonical MHD
paradigm of PWNe \citep{Rees_Gunn74a,Kennel_Coroniti84a,Kennel_Coroniti84b,Begelman_Li92a,Komissarov_Lyubarsky04a,Del-Zanna_Amato+04a,Bogovalov_Chechetkin+05a,Volpi_Del-Zanna+08a,Porth_Komissarov+14a,Olmi_Del-Zanna+16a,Del-Zanna_Olmi17a}. BSPWNe offer the unique opportunity to study the
characteristic of pulsar injection in an environment where  the
typical flow timescales are short, compared to synchrotron cooling
times, even in radio, such that one can safely assume that they trace the
instantaneous injection conditions of the pulsar, as opposed to younger PWNe inside
SNRs, where the contribution to the emission by particles injected
during the entire lifetime of the system makes it harder to
disentagle secular effects from instantaneous properties.

Unfortunately, once the structure of magnetized winds is included, the
dynamics in BSPWNe will turn out to depend on a variety of
different parameters: the anisotropy of the pulsar wind energy flux \citep{Bandiera93a,Wilkin00a,Vigelius_Melatos+07a},
the inclination of the spin axis with respect to the pulsar kick
velocity \citep{Johnston_Hobbs+05a,Ng_Romani07a,Johnston_Kramer+07a,Noutsos_Kramer+12a,Noutsos_Schnitzeler+13a}, the magnetization of the pulsar wind \citep{Bucciantini_Amato+05a}, the presence and fate of a striped
wind region \citep{Mochol17a,Cerutti_Philippov17a,Porth_Komissarov+14a,Komissarov13a,Petri_Lyubarsky07a,Del-Zanna_Amato+04a}, not to mention the conditions in the
ISM, like its neutral fraction \citep{Bucciantini02c,Morlino_Lyutikov+15a}. On top of this, if one wishes to model
the observed properties, also the orientation of the observer's
direction needs to be taken into account, as well as the acceleration
properties of the emitting particles \citep{Olmi_Del-Zanna+14a,Olmi_Del-Zanna+15a,Sironi_Spitkovsky11a,Sironi_Spitkovsky09a}. All of this makes a
complete sampling of the possible parameter space, via brute force
numerical simulations in the proper relativistic MHD regime, quite
challenging. However, if one gives up capturing all the complexity of
the dynamics of the flow, and retains only those key features that are
supposed to characterize the average properties of the system, it is
then possible to develop extremely simplified models, that can at
least try to address the 3D structure of the magnetic field, and the
expected non-thermal emission properties. These might serve both as a
benchmark, with respect to full fledged numerical models, and as a
starting point, to select those configurations that appear to be more
interesting in the light of specific observables.

In this work we develop a formalism to transport the 3D 
magnetic field on a given stationary hydrodynamical configuration,
 and we apply it to the development of models of the magnetic field geometry
for BSPWNe, which we then use to build emission maps, including
polarization, in orded to understand the typical patterns one might
observe in synchrotron emission. 

This work is structured as follows: in Sect.~\ref{sec:magev} we
introduce and describe the formalism we adopt to solve  the induction
equation for
magnetic field evolution; in Sect.~\ref{sec:flow}, we present the
simplified flow structure that we have adopted for BSPWNe; in
Sect~\ref{sec:magstr} we detail how to relate the magnetic field structure 
 in the nebula to the one in the pulsar wind, which is our injection site; in Sect~\ref{sec:res}
 we present and discuss our results. Finally we conclude in Sect.~\ref{sec:conc}.

\section{Transport of a Solenoidal Vector Field}
\label{sec:magev}

The equation that describes the evolution of the magnetic field
$\boldsymbol{B}$, measured in the observer frame, in
Ideal MHD is the induction equation:
\begin{align}
\frac{\partial \boldsymbol{B}}{\partial t}  = -\nabla\times
\boldsymbol{E} = \nabla \times [\boldsymbol{v}\times \boldsymbol{B} ] /c
\end{align}
where $\boldsymbol{E}$ is the electric field in the observer frame
and  $\boldsymbol{v}$ is the flow speed. Together with the solenoidal
condition, $\nabla \cdot \boldsymbol{B} =0$, this equation can be
written as:
\begin{align}
\frac{D \boldsymbol{B}}{D t}  =\frac{\partial \boldsymbol{B}}{\partial
  t} +(\boldsymbol{v}\cdot\nabla) \boldsymbol{B}  = (\boldsymbol{B} \cdot \nabla) \boldsymbol{v} - (\nabla\cdot \boldsymbol{v} ) \boldsymbol{B}\label{eq:lagrangind}
\end{align}
where $D/Dt$ represents the Lagrangian derivative of the field along
the trajectory of a fluid element. This equation allows one to evolve
the magnetic field along the flow, once the velocity field and its
derivatives are known, and it is commonly adopted in Lagrangian
schemes  \citep{Price_Monaghan04a,Rosswog_Price07a,Price11a}. This equation however is not suitable when the
velocity field develops strong discontinuities, like shocks, where the
flow velocity is not properly defined, and local derivatives diverge. One can solve this problem
either introducing some viscosity, and smoothing the shock jumps, or
resorting to an integral version for the evolution of the magnetic
field, which is the approach we adopt in this work.

\begin{figure}
	\centering
	\includegraphics[width=.5\textwidth]{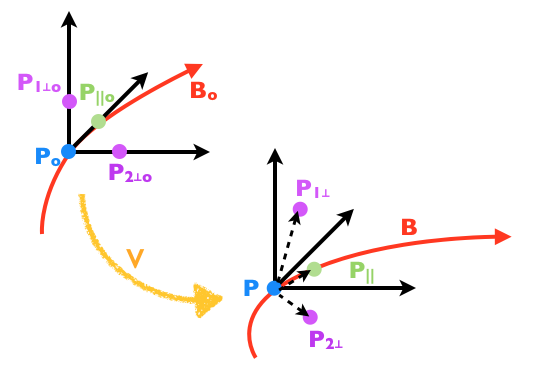}\\
	\caption{Schematic evolution of a fluid elements and its
          nearby fluid elements displaced either in the parallel or
          perpendicular
          directions. $\boldsymbol{P}=\boldsymbol{P}(t,\boldsymbol{P}_{o})$,
 $\boldsymbol{P}_\parallel=\boldsymbol{P}(t,\boldsymbol{P}_{\parallel o})$, 
$\boldsymbol{P}_{1\perp}=\boldsymbol{P}(t,\boldsymbol{P}_{1\perp
            o})$,
          $\boldsymbol{P}_{2\perp}=\boldsymbol{P}(t,\boldsymbol{P}_{2\perp
            o})$, and $\boldsymbol{B}_o=\boldsymbol{B}(0,\boldsymbol{P}_o)$
     }
	\label{fig:transp}
\end{figure}

The flux-freezing condition of Ideal MHD relates the evolution of the
magnetic field to the way nearby fluid elements move with respect to
one another, and in particular to the way their relative displacements
evolve (see Fig.~\ref{fig:transp}).
Let us call $\boldsymbol{P}(t,\boldsymbol{P}_o)$ the trajectory of a
fluid element (the position of a fluid element that at $t=0$ was
located in  $\boldsymbol{P}_o$), and
$\boldsymbol{B}(t,\boldsymbol{P}_o)$ the magnetic
field along the trajectory $\boldsymbol{P}(t,\boldsymbol{P}_o) $ of
that same fluid element at time $t$. 

Let us consider a point $\boldsymbol{P} _{\parallel\;
0} = \boldsymbol{P}_o +\epsilon
\boldsymbol{e}_B(0,\boldsymbol{P}_o)$, where
$\boldsymbol{e}_B(0,\boldsymbol{P}_o)$ is a unitary vector parallel
to the direction of the magnetic field at position $\boldsymbol{P}_o$
and time $t=0$. In the limit  $\epsilon\rightarrow 0$, $\boldsymbol{P} _{\parallel\;
0}$ represents a fluid element displaced by an amount $\epsilon$ from
$\boldsymbol{P}_o$ along the local magnetic
field. Ideal MHD ensures that the two fluid elements remain on the
same magnetic field line. Then, to first order in $\epsilon$, the direction of the magnetic field in
$\boldsymbol{P}(t,\boldsymbol{P}_o)$ will be simply given by:
\begin{align}
\boldsymbol{e}_B(t,\boldsymbol{P}_o)=\frac{
  \boldsymbol{P}(t,\boldsymbol{P}_{\parallel\;o})- \boldsymbol{P}(t,\boldsymbol{P}_o)}{| \boldsymbol{P}(t,\boldsymbol{P}_{\parallel\;o})- \boldsymbol{P}(t,\boldsymbol{P}_o)|}
\end{align}
One can even define a second order accurate estimate of the magnetic
field direction, by taking also a negative displacement by
$-\epsilon$, and using a centered definition of the positional derivative:  
\begin{align}
\boldsymbol{e}_B(t,\boldsymbol{P}) =\frac{
  \boldsymbol{P}(t,\boldsymbol{P}_{+\parallel\;o})- \boldsymbol{P}(t,\boldsymbol{P}_{-\parallel\;o})}{| \boldsymbol{P}(t,\boldsymbol{P}_{+\parallel\;o})- \boldsymbol{P}(t,\boldsymbol{P}_{-\parallel\;o})|}
\end{align}
where $\pm\parallel$ refers to an initial displacement of $\pm
\epsilon$ respectively. Given an initial magnetic field loop,
this equation allows one to trace the evolution of the loop in time,
and to trace the related magnetic surface. The values of $\epsilon$
must be chosen in order to preserve in time the accuracy of the estimate. For
example, it is well known that the displacement will diverge
exponentially in a turbulent flow \citep{Yamada_Ohkitani87a,Biferale_Boffetta+05a,Yamada_Ohkitani98a,Salazar_Collins09a,Berera_Ho18a}, while in a laminar flow
it will likely remain bound.

On the other hand, in Ideal MHD, the evolution of the strength of the magnetic field
$\boldsymbol{B}(t,\boldsymbol{P}_o)$
along the fluid element trajectory $\boldsymbol{P}(t,\boldsymbol{P}_o)$
is related to the evolution of the cross section of an infinitesimal
magnetic flux tube centered on the same fluid element,
$\Phi(t,\boldsymbol{P}_o)$:
\begin{align}
\frac{|\boldsymbol{B}(t,\boldsymbol{P}_o)|}{|\boldsymbol{B}(0,\boldsymbol{P}_o)|} =\frac{\Phi(0,\boldsymbol{P}_o)}{\Phi(t,\boldsymbol{P}_o)}
\end{align}
To trace the evolution of the cross section of an infinitesimal
magnetic flux tube, one can follow the evolution of two fluid elements
originally (at $t=0$) displaced in directions perpendicular to the
magnetic field. Let $\boldsymbol{e}_{1\perp}$ and $
\boldsymbol{e}_{2\perp}$ be unitary vectors such
that:
\begin{align}
\boldsymbol{e}_{1\perp}\cdot
\boldsymbol{e}_{2\perp}=0, \quad
\boldsymbol{e}_{1\perp}\cdot
\boldsymbol{e}_{B}(0,\boldsymbol{P}_o) = 0, \quad \boldsymbol{e}_{2\perp} \cdot
\boldsymbol{e}_{B}(0,\boldsymbol{P}_o) = 0,\!
\end{align}
then one can define two fluid elements originally displaced
perpendicular to the local magnetic field as:
\begin{align}
\boldsymbol{P} _{1\perp\;
o} = \boldsymbol{P}_o +\epsilon
\boldsymbol{e}_{1\perp}\quad
\boldsymbol{P} _{2\perp\;
o} = \boldsymbol{P}_o +\epsilon
\boldsymbol{e}_{2\perp}
\end{align}
and the initial (at $t=0$) infinitesimal cross section
of the flux tube at $\boldsymbol{P}_o$ as: 
\begin{align}
\Phi(0,\boldsymbol{P}_o) = |(\boldsymbol{P} _{1\perp\;
o}-\boldsymbol{P} _{
o})\times (\boldsymbol{P} _{2\perp\;
o}-\boldsymbol{P} _{
o})|=\epsilon^2
\end{align}
In general, depending on the shear of the velocity field, the evolution of fluid elements originally  displaced
perpendicular to the magnetic field, will not keep them
perpendicularly displaced. One however can redefine a perpendicular
projection as:
\begin{align}
\boldsymbol{\xi}_1&=\boldsymbol{P}(t,\boldsymbol{P}_{1\perp\;o})-\boldsymbol{P}(t,\boldsymbol{P}_{o})+\nonumber\\
&\quad\quad-[(\boldsymbol{P}(t,\boldsymbol{P}_{1\perp\;o})-\boldsymbol{P}(t,\boldsymbol{P}_{o}))\cdot \boldsymbol{e}_B(t,\boldsymbol{P}_o)]\boldsymbol{e}_B(t,\boldsymbol{P}_o)
\end{align}
and similarly for $\boldsymbol{\xi}_2$. Then the infinitesimal flux
tube cross section will be given by $\Phi(t,\boldsymbol{P}_o) =
|\boldsymbol{\xi}_1\times \boldsymbol{\xi}_2 |$. Again it is possible
to define a second order estimate for the evolution of the cross
section of the flux tube, using displacements in both the positive and
negative direction, and taking central differences in the definition of
the $\boldsymbol{\xi}$ projections. 

Given that the evolution of the trajectory of a fluid element, and of
nearby displaced fluid elements, is simply given by the velocity
field, and does not involve its derivatives, one can use this approach
to trace the evolution of the magnetic field in the presence of
discontinuities in the velocity as those expected in the case of
shocks. Note moreover that this approach makes no assumption on the
velocity field, excluding only those highly turbulent flows where
nearby fluid elements rapidly drift apart. We have verified that on
smooth differentiable flow fields this approach, based on
displacments, gives equivalent  results to direct integration of Eq.~\ref{eq:lagrangind}.

\section{A Laminar Flow Model}
\label{sec:flow}

\begin{figure}
	\centering
	\includegraphics[width=.45\textwidth]{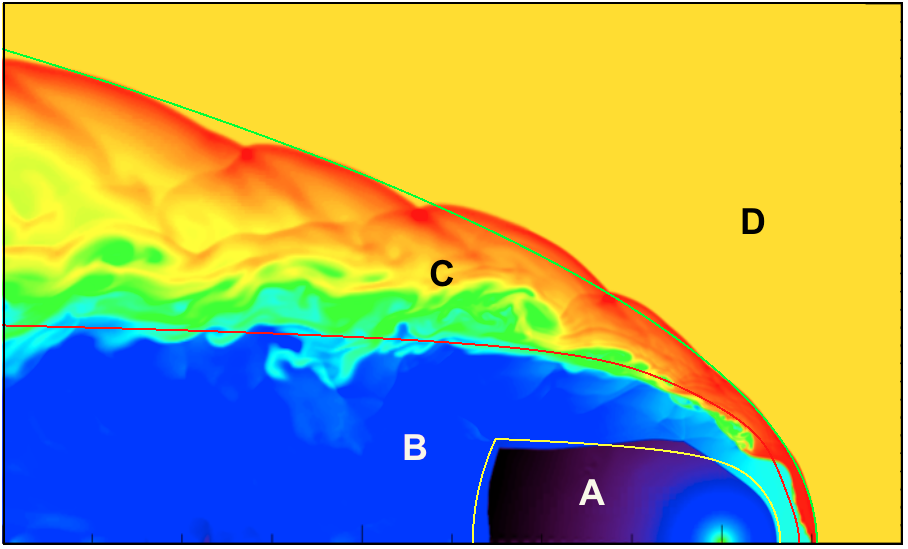}\\
	\includegraphics[width=.45\textwidth]{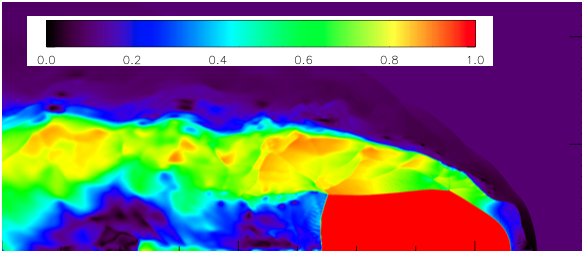}\\
        \includegraphics[width=.5\textwidth]{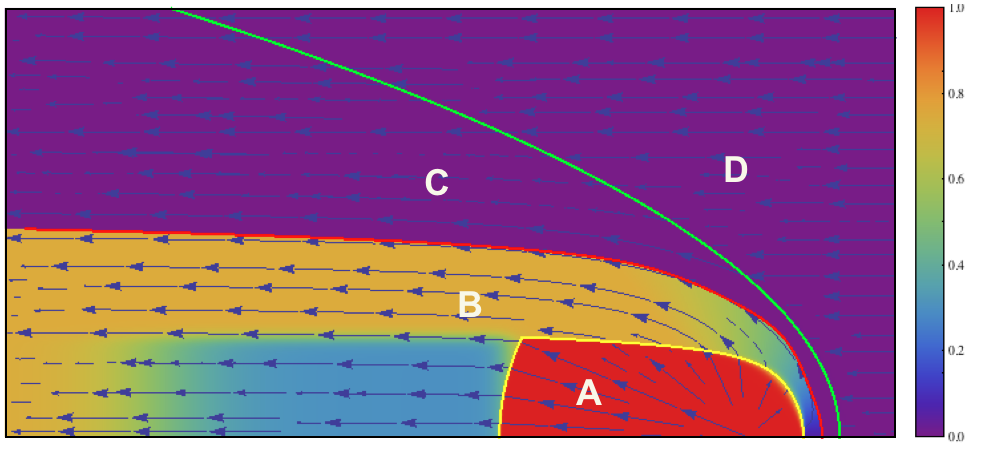}\\
	\caption{Upper panel: the color image represents the density
          structure of a bow-shock taken from an axisymmetric
          relativistic hydrodynamical simulation. The superimposed
          curves are the location of the termination shock (yellow),
          the contact discontinuity (red) and the forward
          bowshock (green) of our analytical steady state
          model.  Middle panel: the color image represents the velocity magnitude
           of the bow-shock taken from the same axisymmetric
          relativistic hydrodynamical simulation, at the same time. 
        Lower panel: the flow structure assumed in out
          model. The color map represents the magnitude of the velocity
          field, while the arrows indicate its direction. The colored
          curves are the same as in the upper panel. Labels refer to:
          A unshocked pulsar wind, B shocked pulsar wind, C shocked
          ISM material, D unshocked ISM.
     }
	\label{fig:vel}
\end{figure}

In this section we describe the velocity field we used to model the
transport of magnetic field in BSPWNe. We begin
by briefly reviewing the key characteristics of BSPWNe. Let us place
ourself in a reference frame centered on the PSR, with the
$z-$axis aligned with the pulsar speed. In this reference frame the
ISM is seen moving in the negative $z-$direction. Given the typical
values of the pulsars' velocity $\sim 100-400$ km s$^{-1}$, and the
typical temperatures of the ISM $\sim 10^4-10^5$ K, such flow will be
highly supersonic. The interaction of
the relativistic pulsar wind with the incoming ISM, produces a
cometary-like nebula of shocked pulsar wind and ISM material. Four
regions can be identified in this system. With reference to
Fig.~\ref{fig:vel}, moving from the pulsar outward one finds:
\begin{itemize}
\item In inner region (A), containing
the relativistic, magnetized and cold pulsar wind, which moves with high Lorentz
factor in the radial direction. 
\item A cometary-like region (B) of pulsar wind material, which has been shocked,
  heated and slowed down to sub-relativistic speeds in a bullet shaped
  termination shock (TS). It is this region that is seen in non-thermal
  radio and X-rays  synchrotron emission. 
\item A region (C) of ISM material shocked in a forward bow-shock, and
  separated from the former region B by a contact discontinuity (CD). This is
  the region usually observed in H$_\alpha$ emission. 
\item An outer region (D) occupied by the supersonic ISM flow. 
\end{itemize}
Each one of these regions is separated from the other by sharp
transitions, either a shock or a CD. Due to the balance between the ram
pressure of the pulsar wind and the one of the incoming ISM, material
in region B is forced to flow into a collimated tail. While in the
head, the flow speed of region B is strongly subsonic, and just a
small fraction of the speed of light, as it moves sideways toward the
tail, it accelerates, becoming supersonic and reaching a speed as high
as $0.7-0.8c$. Downstream of the Mach Disk that forms the back side of
the wind termination shock, a low velocity inner channel develops. It
is found in numerical simulations that this low velocity inner
channel can extend up to twice the distance of the Mach Disk, where,
due to the interaction with the faster surrounding flow, it finally
accelerates to supersonic speeds. On the other hand typical flow
velocities in region C are much smaller (of the order of a typical
pulsar speed in the ISM). This leads to the development of a strong
shear layer at the CD. In principle the CD is highly prone to
the development of Kelvin-Helmholtz instability, which might mix ISM
and pulsar wind material. This instability was not observed in earlier
numerical simulations of these systems \citep{Bucciantini02a,Gaensler_van-der-Swaluw+04a}, due to low resolution. It is
not clear if the presence of a strong magnetic field can stabilize
the CD, or if it introduces further channels for instabilities. Magnetized
simulations carried in the fully axisymmetric regime \citep{Bucciantini_Amato+05a}, where the
stabilizing effect of magnetic tension is absent, cannot be used to
address this issue. In principle the growth of instabilities at the CD
can lead to the formation of internal waves/shocks that can also affect the shape of the termination and forward
shocks, making these systems highly dynamical. Unfortunately, despite
the fact that such variability is expected to have typical timescales of the
order of the bow-shock light crossing time, no repeated follow up
campaign for these systems has yet been carried out, to assess this problem.

However,  we are here interested in developing a simplified
model, capable of  capturing the key properties of the flow, without
all the complexity that only full fledged 3D high resolution numerical
simulations can handle. We therefore chose to assume a steady state, axisymmetric
laminar flow structure,  representative of typical average flow
conditions, as can be derived from existing simulations. Based on
hydrodynamical 2D axisymmetric numerical simulations we have derived
analytical formulas for the shape of the termination shock and of the
forward bow-shock. The size of the various regions is scaled such that
the CD along the  $z-$direction is located at a unitary distance from the
pulsar. This is equivalent to normalizing the lengths of the model to
the stand-off distance defined as:
\begin{align}
d_{o}=\sqrt{\frac{L}{4\pi c \rho_o V^2}}
\end{align}
where $L$ is the pulsar spin-down luminosity, $\rho_o$ is the ISM
density, and $V$ is the relative speed of the pulsar with respect to
the ISM. The flow is moving radially at the speed of light within the
termination shock (in region A),  $\boldsymbol{v}=
c\boldsymbol{e}_r$, while matter in the outer ISM (region D) is
moving with a speed  $-V$ along the $z-$ axis, $ \boldsymbol{v}=
-V\boldsymbol{e}_z$. To define the direction of the flow field in both
region B and C we adopt the same analytical formula. The only
difference in the two regions is the velocity magnitude. The use of
the same analytical formula for the flow direction, ensures that the
contact discontinuity is mathematically well defined (if one had used
two different prescriptions for the direction of the flow speed in region B and C, then
it would have been necessary to enforce the correct matching at the
supposed CD).  The analytical prescription for the flow field in
region B and C is chosen in order to reproduce at best the typical
location of the CD (see App.~\ref{sec:app}). This (together with the decision to adopt simple
analytical prescriptions) implies that, at the two shocks, the jump
conditions are not satisfied. This however affects the flow only
locally. The magnitude of the velocity in region B and C is chosen
in order to reproduce the typical trend found in numerical
simulations. In region B (the one we are interested in for non-thermal
emission) the velocity is seen to rise linearly, from very small
values in the head, to values $\sim 0.75-0.85c$ in the tail. We therefore assume a
linear trend with angular displacement from the
$z-$axis. We also include the presence a slowly moving inner channel
downstream of the Mach Disk. Fig.~\ref{fig:vel} illustrates the velocity field and
magnitude we have adopted, to be compared with existing numerical
simulations (see for example Fig.~1 of \citet{Bucciantini_Amato+05a}). Despite its
simplicity the flow field encapsulates all the key properties found in
more sophisticated 2D axisymmetric simulations.  In our simplified
  model we assume that the energy flux in the pulsar wind is
  isotropic. However numerical simulations of pulsar magnetospheres
  \citep{Spitkovsky06a,Tchekhovskoy_Spitkovsky+13a,Tchekhovskoy_Philippov+16a}
  show that the wind has likely a higher equatorial energy
  flux. This has consequences on the shape of the TS, which partly reflects
  the energy flux of the wind \citep{Del-Zanna_Amato+04a}, and on the
  post shock flow, creating flow channels that can trigger the
  developement of turbulence and affect the overall variability of the
  system \citep{Camus_Komissarov+09a}.

\section{Magnetic Field Injection}
\label{sec:magstr} 

\begin{figure}
	\centering
	\includegraphics[width=.5\textwidth]{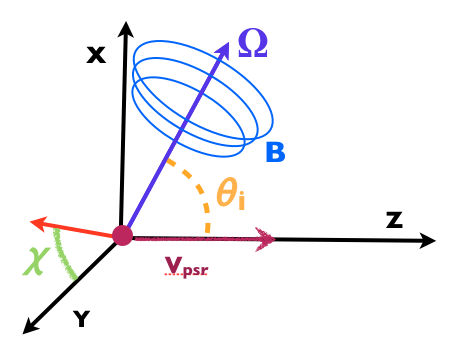}\\
	\caption{Geometry of the magnetic field in the pulsar
          wind. The $z-$axis is aligned to the pulsar speed $\boldsymbol{v}_{\rm
            psr}$ with respect to the ISM (or equivalently the ISM
          speed in the reference frame of the pulsar). $\boldsymbol{\Omega}$ is the
          pulsar spin axis, around which the pulsar wind magnetic
          field $\boldsymbol{B}$ is wound-up, which lays in the $x-z$
          plane and is inclined $\thi$ with respect to the $z-$
          axis. $\chi$ indicates the inclination angle of the
          direction pointing toward the observer, which is assumed to be
          perpendicular to the $z-$axis (the pulsar velocity is in the
          plane of the sky).
     }
	\label{fig:geom}
\end{figure}

The use of a laminar steady-state flow field ensures that each point
of the BSPWN system can be associated one to one to an injection
location. For the ISM the injection can be placed anywhere in region
D. For the pulsar material injection can be placed at any
radius in region A, where one can use standard predictions for the
structure and geometry of the magnetic field in pulsar winds. In
particular we set our injection boundary for the pulsar magnetic field
on a sphere with radius $R_{o} = 0.5d_o$. We are interested in modeling
the non-thermal emission from region B so we assume the ISM material
to be unmagnetized. In general the pulsar spin axis will be inclined
by and angle $\thi$ with respect to the $z-$axis, as shown in
Fig.~\ref{fig:geom}. The magnetic field in the pulsar wind is assumed to be
purely toroidal, and symmetric with respect to the spin axis. In
principle the magnetic field strength will depend on the latitude $\psi$
with respect to the spin axis. Many existing numerical models for
example assume a sin-like dependence
\citep{Komissarov_Lyubarsky04a,Volpi_Del-Zanna+08b,Porth_Komissarov+14a,Olmi_Del-Zanna+15a,Olmi_Del-Zanna+16a},
and possibly the presence of en equatorial unmagnetized region, to
account for dissipation of a striped wind.

In a cartesian reference frame centered on the pulsar with the $x-$
axis chosen such that the pulsar spin axis lays on the $x-z$ plane (Fig.~\ref{fig:geom}),
the magnetic field $\boldsymbol{B}_o$ at a point
$\boldsymbol{P}_o = [x_o,y_o,z_o]$ in the wind will be given, in terms
of cartesian components,
by:
\begin{align}
\boldsymbol{B}_o \propto \frac{F(\psi)}{r_o\mathcal{R}}\begin{cases}
 y_o \cos{\thi}\boldsymbol{e}_x\\
 (z_o \sin{\thi}-x_o\cos{\thi}) \boldsymbol{e}_y\\
 -y_o\sin{\thi}\boldsymbol{e}_z
\end{cases}
\end{align}
where $r_o=\sqrt{x_o^2+y_o^2+z_o^2}$, $\mathcal{R}^2 =
y_o^2+x_o^2\cos^2{\thi}+z_o^2\sin^2{\thi}-x_oz_o\sin{2\thi}$. $\psi$
represents the angular distance of a point from the pulsar spin axis,  $\cos{\psi}=[x_o
\cos{\thi}+z_o\sin{\thi}]/r_o$, and $F(\psi)$ parametrizes the
latitudinal dependence of the magnetic field strength.

\section{Results}
\label{sec:res}

We present here the results of a serie of models done using various
inclinations of the pulsar spin axis $\thi$ and various orientation
$\chi$ of the observer. We are mostly interested in modeling the
magnetic field structure and simulating the non-thermal radio
emission, including polarization, in oder to obtain an understanding of
the basic features one might expect to see in these systems. In the
following, in line with the standard assumptions on the magnetic field
distribution in pulsar winds, we assume $F(\psi) =\sin{\psi}$, unless
otherwise stated. Moreover we assume that the velocity field is
independent on the magnetic field strength. This is formally correct only as
long as  the magnetic field energy is sub-equipartition. 

\subsection{Magnetic surfaces}

\begin{figure}
	\centering
	\includegraphics[width=.45\textwidth]{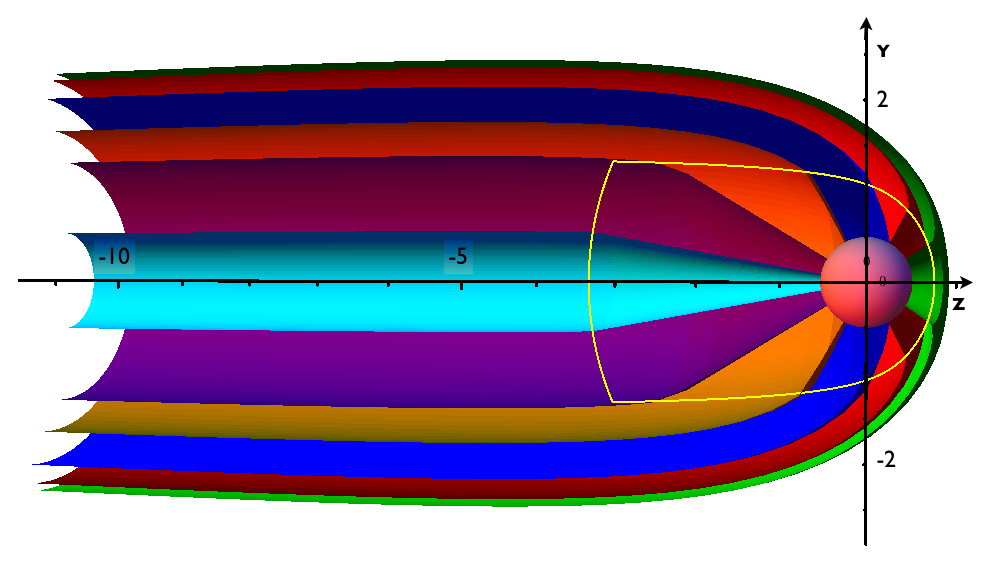}\\
	\includegraphics[width=.45\textwidth]{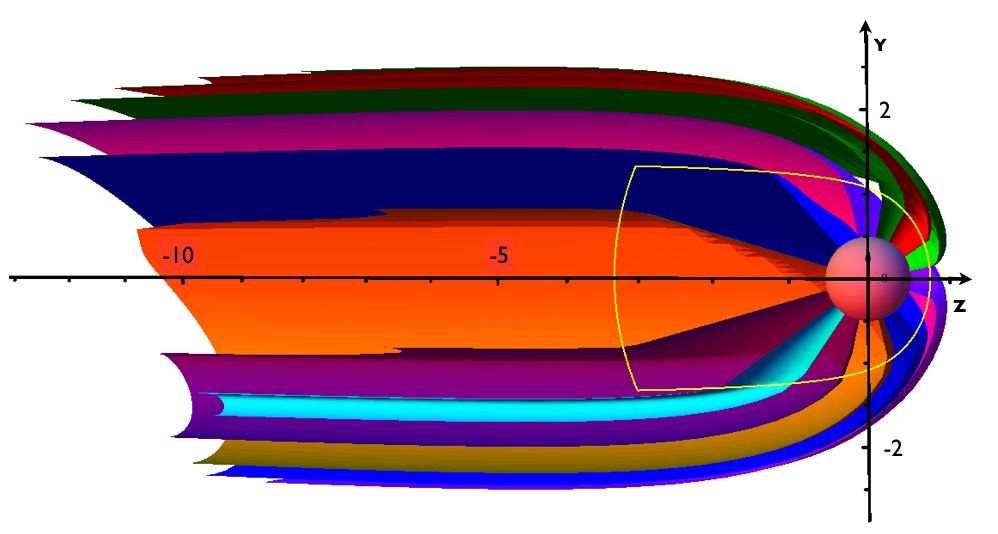}\\
        \includegraphics[width=.45\textwidth]{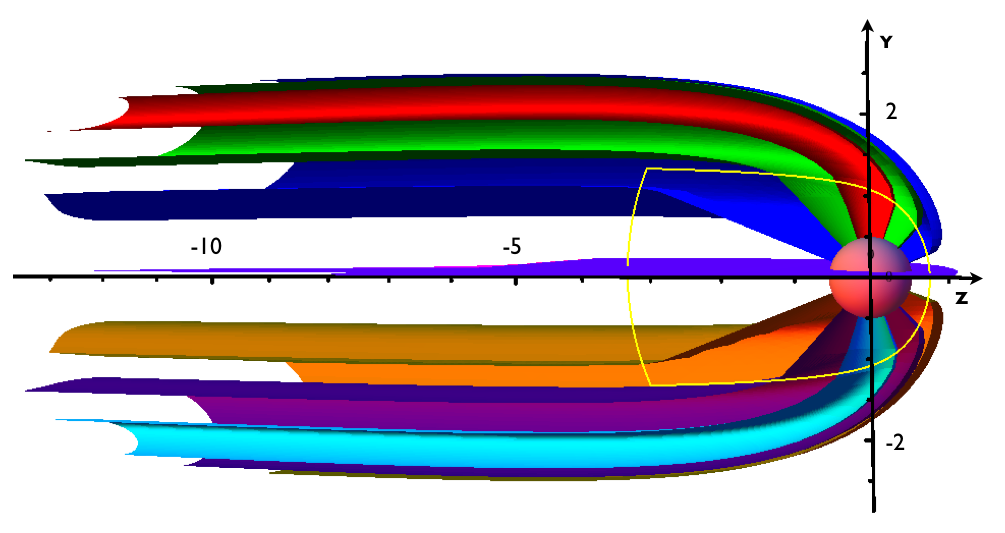}\\
	\caption{Sections of magnetic field surfaces, for various inclination of
          the pulsar spin axis. From top to bottom
          $\thi=0,45^\circ,90^\circ$. Magnetic surfaces have been
          selected to sample as best as possible the distribution of
          magnetic loops in the nebula. The yellow curve is a section
          of the pulsar wind TS.
     }
	\label{fig:magsurf}
\end{figure}

In a steady state laminar flow, magnetic loops, coming from the wind,
will retain their coherence and trace, during their evolution, the
location of the magnetic surface to which they belong. A set of nested
magnetic surfaces will then arise as a consequence of the flow. In
Fig.~\ref{fig:magsurf} we show the shape of magnetic field surfaces, for
various inclinations of the pulsar spin axis. In the co-aligned case,
$\thi=0$, the magnetic surfaces are axisymmetric, while for other
inclinations, a {\it magnetic chimney} forms on the front side, with
magnetic loops strongly stretched, and piled-up. It is evident that the flow
structure tends to concentrate the magnetic surfaces toward the
contact discontinuity. This effect was already observed in earlier
axisymmetric simulations  \citep{Bucciantini_Amato+05a}. This will lead to the formation of a
magnetopause, where the magnetic field strength can rise above
equipartition. Magnetopauses are a common feature in magnetically
confined flows, and are sites of possible violent reconnection and
dissipation \citep{Mozer_Torbert+78a,Galeev83a,Alexeev_Kalegaev95a}. In the axisymmetric case, the
structure of the magnetic surfaces ensures that there will be no
polarity inversion close to the CD, suppressing in principle
reconnection events. On the contrary, in the magnetic
chimney that form for other values of $\thi$, field lines of different
polarity are brought close together, in principle creating the conditions
for fast reconnection and dissipation. It is then possible that the
magnetic chimney behaves as a highly dissipative structure. On the
other hand, the magnetic surfaces are quite open on the back side,
downstream of the Mach Disk, suggesting that low values of the
magnetic field might be characteristic of the inner flow channel.
Indeed the backward magnetic chimney shows no sign of stretching or compression.

\subsection{Emission maps}
Let us recall that we are interested in radio emission, and 
radio emitting particles have synchrotron lifetime longer that
the flow time in the nebula. Following \citet{Del-Zanna_Volpi+06a}, if we assume that the emitting pairs are distributed in energy
$\epsilon$, according to a power-law 
 $n(\epsilon)=K \epsilon^{-(2\alpha+1)}$, then
 their local synchrotron emissivity at frequency $\nu$, toward the observer will be:
\begin{align}
j(\nu,\boldsymbol{n})= C |\boldsymbol{B}'\times
\boldsymbol{n}'|^{\alpha+1} D^{\alpha+2} \nu^{-\alpha}\label{eq:spectral}
\end{align}
where $\boldsymbol{B}'$ and $\boldsymbol{n}'$ are the magnetic field
and the observer direction measured in the frame comoving with the
flow, and $C$ is a normalization constant dependent on $K$. $D$ is the Doppler boosting coefficient
\begin{align}
D=\frac{\sqrt{1-\beta^2}}{1-\boldsymbol{\beta}\cdot \boldsymbol{n} }=
  \frac{1}{\gamma (1-\boldsymbol{\beta}\cdot \boldsymbol{n})}
\end{align}
where $\gamma$ is the Lorentz factor of the flow,  $\boldsymbol{\beta}$ and $\boldsymbol{n}$ respectively the flow speed normalized
to $c$ and the observer direction, both measured in the observer
frame. Now, in terms of quantities measured in the observer frame
(unprimed), one has
\begin{align}
 |\boldsymbol{B}'\times
\boldsymbol{n}'|
=\frac{1}{\gamma}\sqrt{B^2-D^2(\boldsymbol{B}\cdot\boldsymbol{n})^2+2\gamma
D (\boldsymbol{B}\cdot\boldsymbol{n})(\boldsymbol{B}\cdot\boldsymbol{\beta})}
\end{align}
One can also compute the polarization angle $\xi$, that enters into
the definition of the Stoke's parameters $Q$ and $U$. Choosing a Cartesian
reference frame with the observer in the $X$ direction, one finds
\begin{align}
\cos{2\xi} =\frac{q_Y^2-q_Z^2}{q_Y^2+q_Z^2},\quad\quad \sin{2\xi} =-\frac{q_Yq_Z}{q_Y^2+q_Z^2},
\end{align}
where
\begin{align}
q_Y=(1-\beta_X)B_Y+\beta_Y B_X\quad\quad q_Z=(1-\beta_X)B_Z+\beta_Z B_X
\end{align}
In principle the local emissivity also depends on the local density of
emitting particles ($K$). This, as well as the power-law index $\alpha$, will in principle be a function of the
injection location along the TS (different locations being
characterized by different
physical conditions in terms of wind magnetization, presence of a
striped component, wind Lorentz factor etc....). In our Lagrangian
formalism, we can trace the injection location along the TS for each
fluid element in the BSPWN. However, given that there is no consensus
on how particles are accelerated at a relativistic termination shock,
that current models provide little hint on possible injection recipes,
and that observations do not have enough resolution to assess
possible variations of the spectral index, 
one can assume that the particle power-law index $\alpha$
is uniform in the nebula. Furthermore, we also assume, for simplicity,
that the normalization constant $K$ is also uniform. This is
equivalent to the assumption of a uniform density for the emitting
particles. Given that the divergence of the flow field we use in our
model is small (at most a factor 2 between the head and the tail),
this is roughly equivalent to the assumption of uniform injection at 
the TS. In all of the following, to limit the possible parameter
space, we consider only the case where the observer's direction is
perpendicular to the pulsar velocity (the pulsar velocity is in the
plnae of the sky). In this case, the observer direction is
parametrized only by the viewing angle $\chi$ (see Fig.~\ref{fig:geom}).

\begin{figure}
	\centering
	\includegraphics[bb=30 90 550 330,width=.45\textwidth,clip]{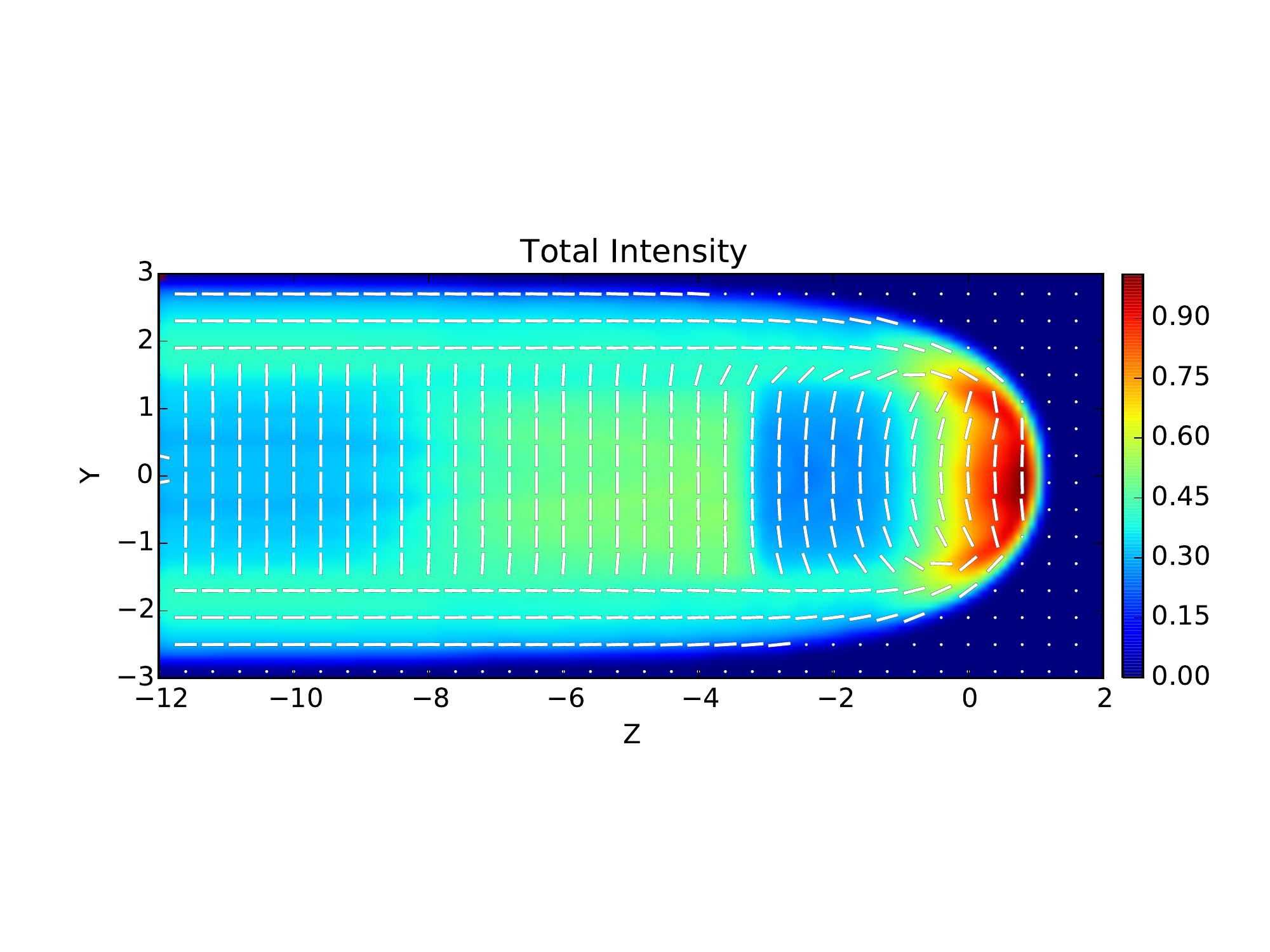}\\
	\includegraphics[bb=30 90 550 330,width=.45\textwidth,clip]{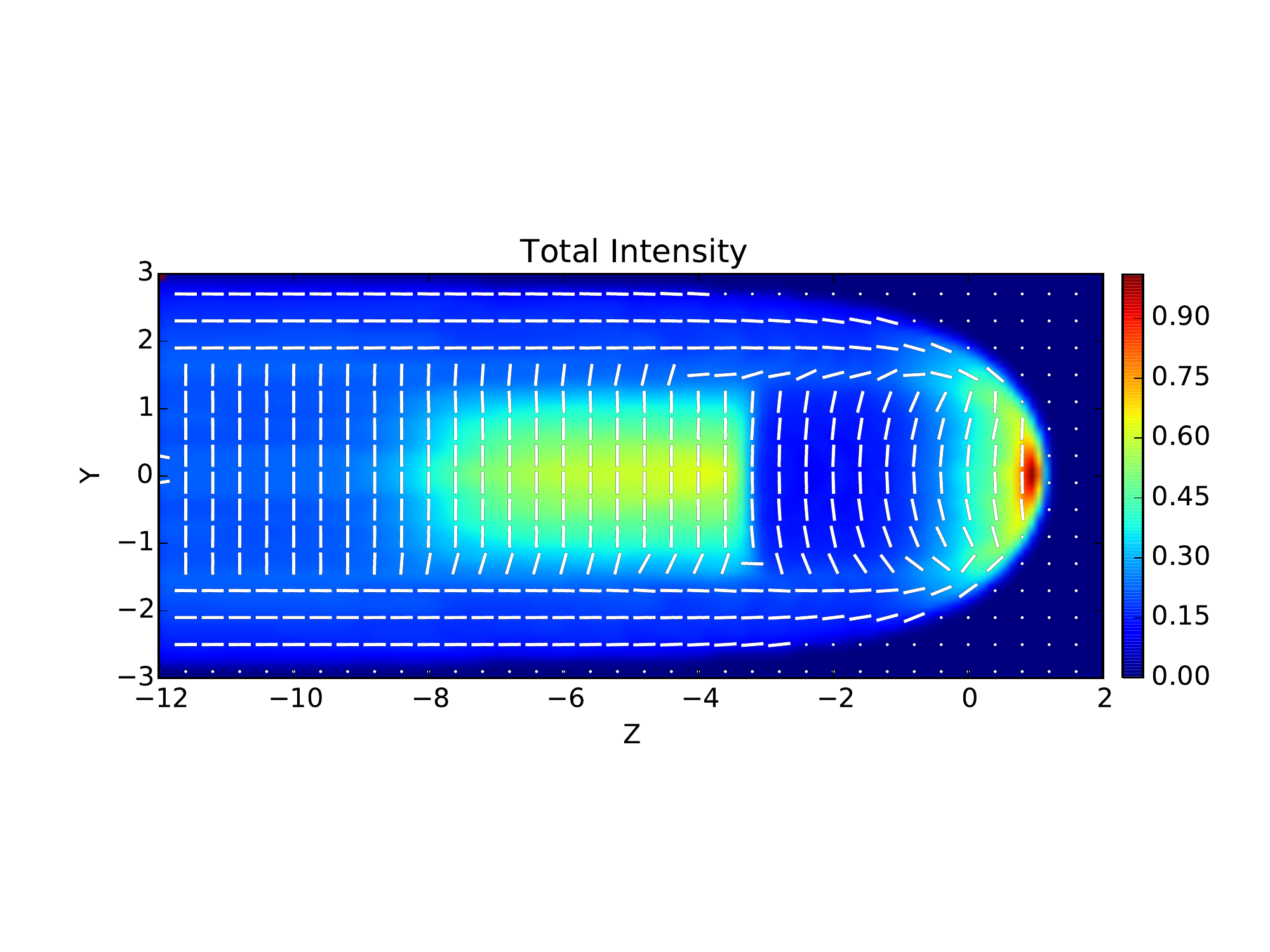}\\
        \includegraphics[bb=30 90 550 330,width=.45\textwidth,clip]{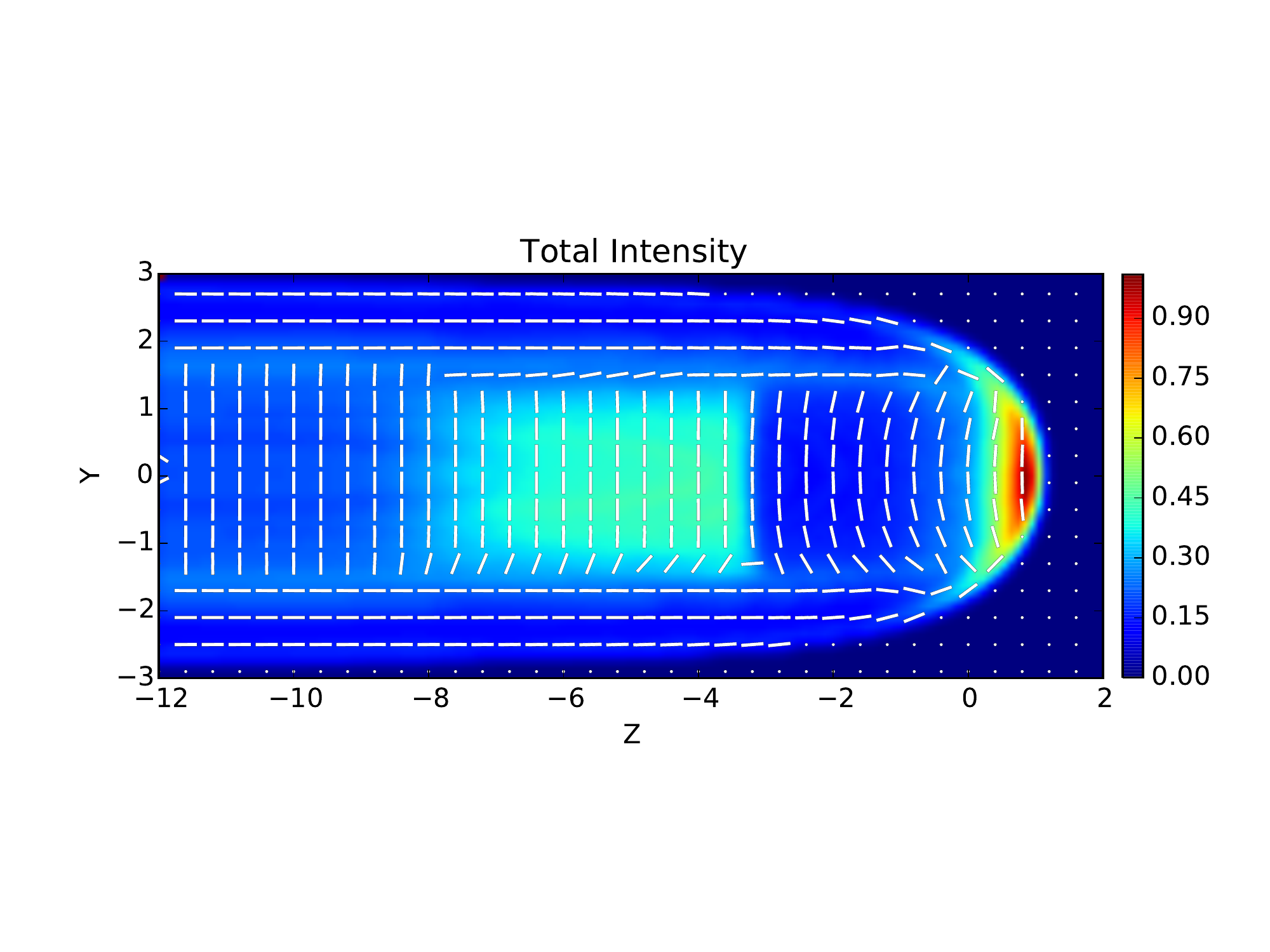}\\
	\caption{Total synchrotron intensity in the completely
          axysymmetric case $\thi=0$ for different choices of the
          magnetic field distribution in the wind: upper panel
          $F(\psi)=\sin{(\psi)}$, middle panel $F(\psi)=const$, lower panel
          $F(\psi)=\sin{(\psi)} \tanh{(\Pi/2-\psi)}$ (representative
          of a large striped wind). Intensity is
          normalized to the maximum. Dashes indicate the orientation of
          the polarization vector (not its amplitude). The pulsar is
          located in $Y=0$ $Z=0$.
     }
	\label{fig:intaxi}
\end{figure}

In the axisymmetric case, $\thi=0^\circ$, the emission pattern is independent
of the viewing angle $\chi$. In Fig.~\ref{fig:intaxi} we show the
expected synchrotron intensity for a flat radio spectrum, $\alpha=0$, and for
three different choices of magnetic field distribution in the
wind.  Given that we are interested in the polarimetric signatures
associated to the magnetic field geometry, we
opted for a flat spectrum in accordance with the typical radio
spectral indexes of PWNe \citep{Gaensler_Slane06a}, and radio
bow-shocks \citep{Predehl_Kulkarni95a,Ng_Gaensler+10a,Ng_Bucciantini+12a}. In
the same figure the polarization pattern is also shown. In all cases
the intensity peaks in the very head. This is due to Doppler boosting,
given that it is the only location in the nebula where the flow
velocity points toward the observer. What distinguishes the various
cases, is how narrow the intensity peak is. Obviously, this depends on
the strength of the magnetic field at the pole, with respect to the
one at the equator (the level of magnetic anisotropy in the wind),
which is maximal for $F(\psi)= \sin{(\psi)}$, and minimal for
$F(\psi)= const$. The other interesting common feature, also connected to
Doppler boosting, is the high brightness region located between $Z=-8$
and $Z=-3$ that corresponds to the slow inner channel that forms
downstream of the Mach Disk. Being the velocity in this channel
smaller than in the surrounding region, the emission is less
de-boosted. The brightness contrast again depends on the magnetic
anisotropy, being maximal for $F(\psi)= const$. In general, in the case $F(\psi)=
\sin{(\psi)}$, the brightness contrast among the various features, is
so attenuated that on average the surface brightness looks more
uniform, in the volume of the nebula. 

\begin{figure}
	\centering
	\includegraphics[bb=30 90 550 330,width=.45\textwidth,clip]{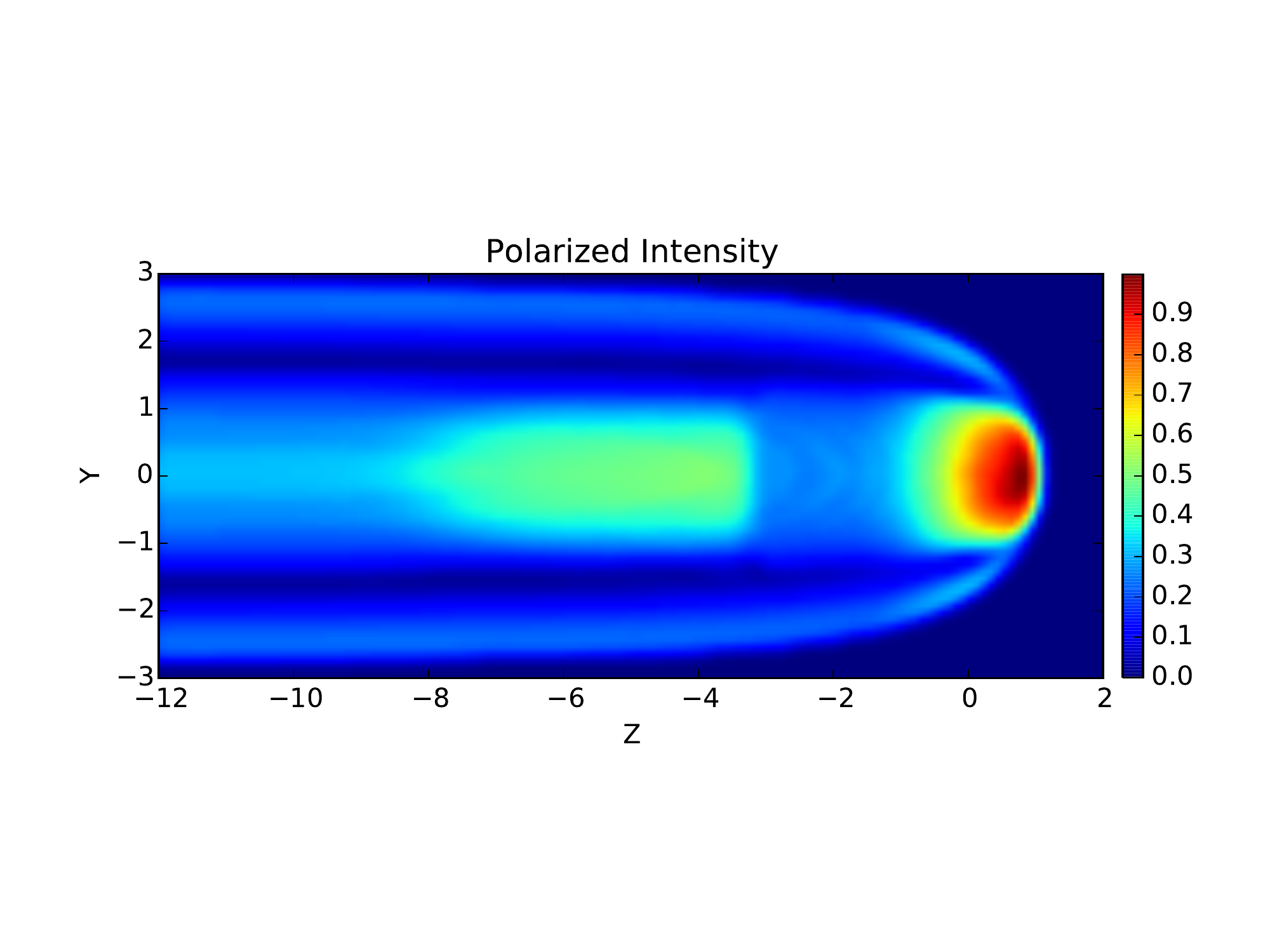}\\
	\includegraphics[bb=30 90 550 330,width=.45\textwidth,clip]{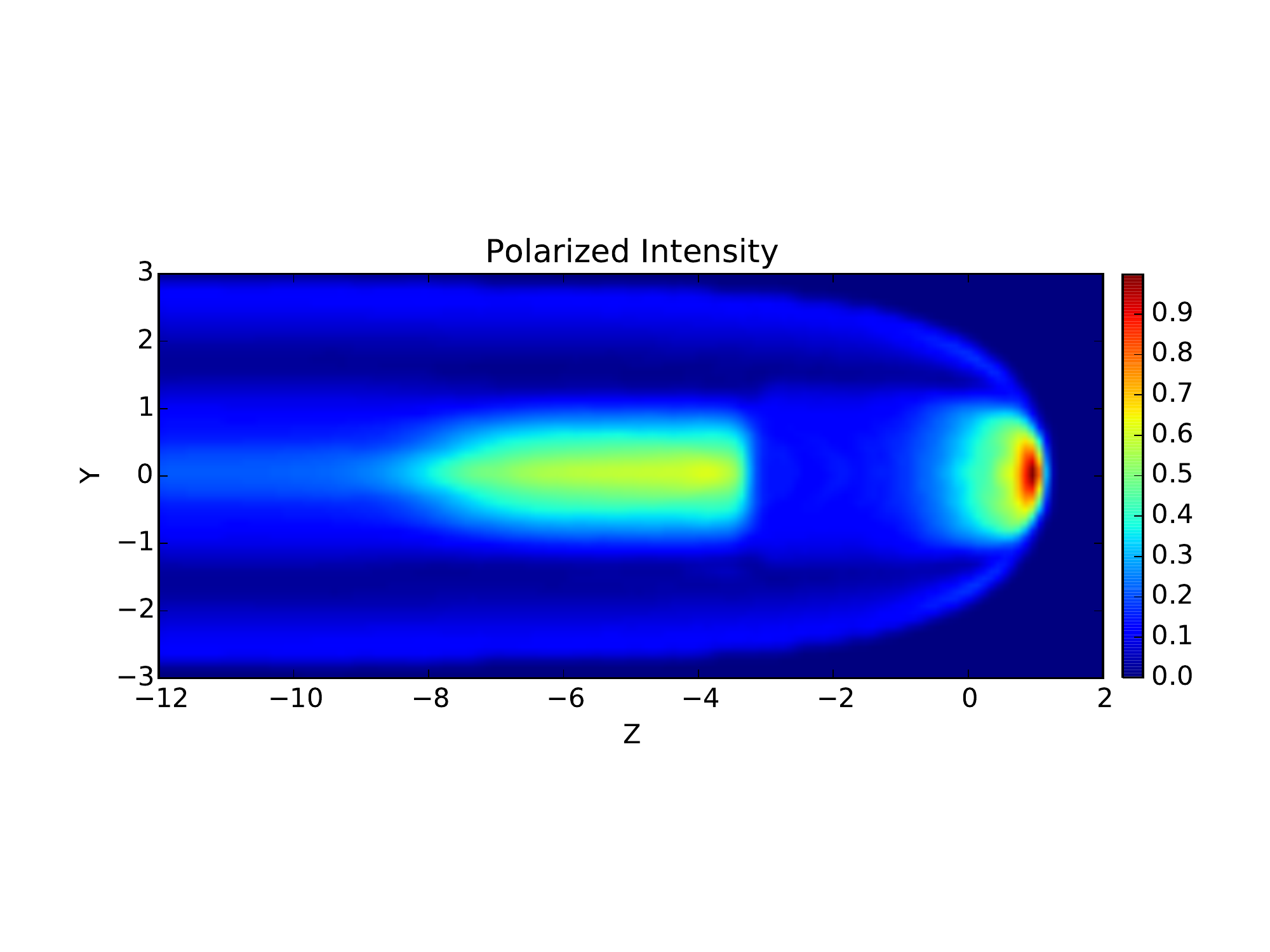}\\
        \includegraphics[bb=30 90 550 330,width=.45\textwidth,clip]{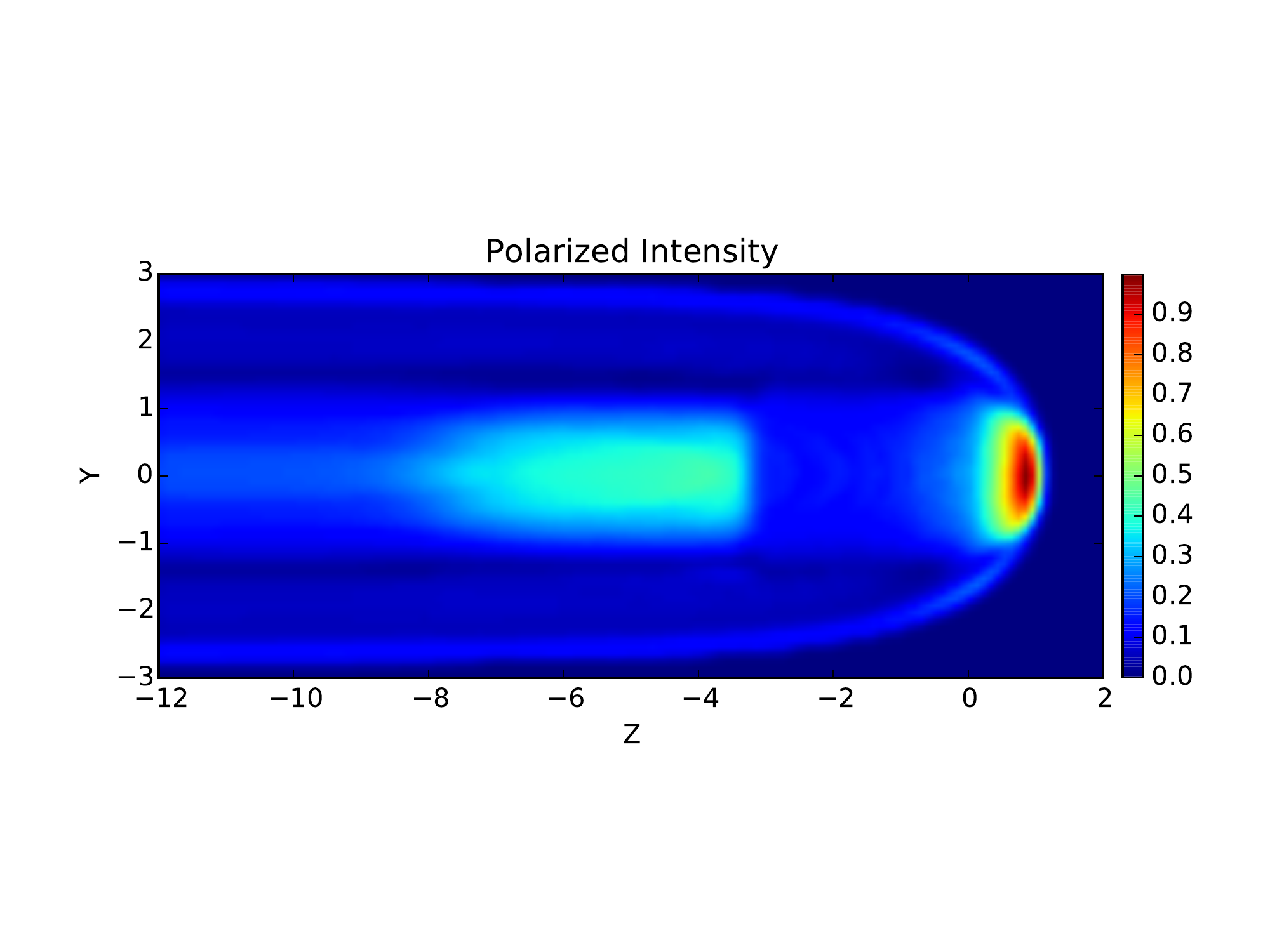}\\
	\caption{Same as Fig.~\ref{fig:intaxi} but for polarized
          intensity. Maps are normalized to the maximum of the total intensity.
     }
	\label{fig:polaxi}
\end{figure}

\begin{figure}
	\centering
	\includegraphics[bb=30 90 550 330,width=.45\textwidth,clip]{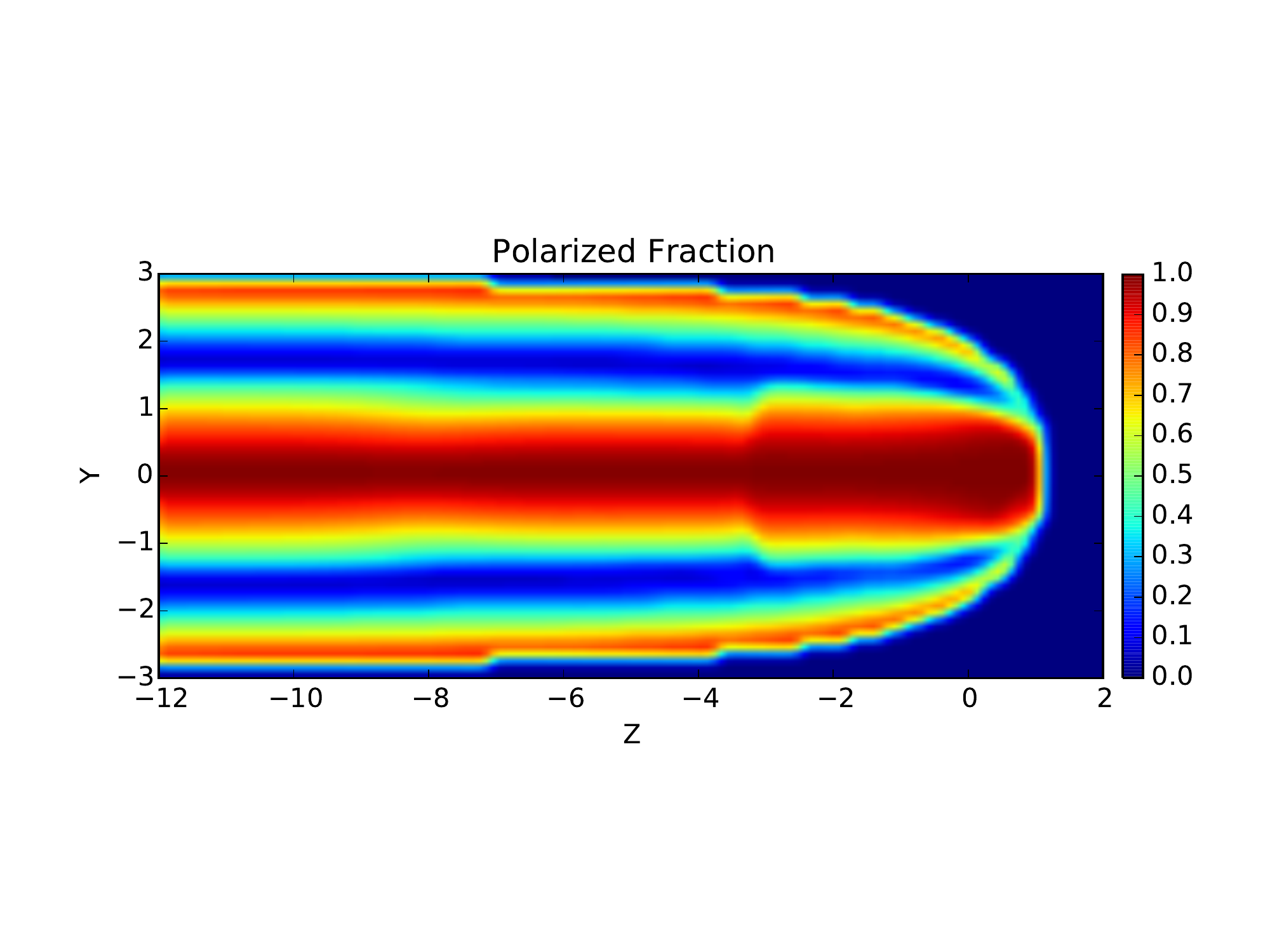}\\
	\caption{Polarized fraction for the case $\thi=0$,  $F(\psi)=\sin{(\psi)}$.
     }
	\label{fig:faxi}
\end{figure}

In the same 
Fig.~\ref{fig:intaxi}, we also show for the same configurations the
polarization angle (orthogonal to the electric field). Given that the
magnetic field is, in the axisymmetric case, always orthogonal to the
pulsar kick velocity, one would have naively expected the polarization
to be everywhere orthogonal to the $Z$-axis. However, it is well known
that relativistic motion leads to polarization angle swing
\citep{Bjornsson82a,Qian_Zhang03a,Bucciantini_del-Zanna+05a,Bucciantini_Bandiera+17a},
and this becomes dominant in the outer fast flow channel. In
Fig.~\ref{fig:polaxi} we show the polarized intensity. The same
considerations discussed previously for the level of brightness
contrast, and its dependence on the magnetic anisotropy of the wind,
apply also for polarized emission. The polarization angle swing leads
to a depolarized region between the inner slow channel and the
CD. Interestingly the inner slow channel is more evident in polarized
light than in total light. The polarized intensity in the inner channel
ranges from 40\% to 60\% of the maximum total intensity, while the bulk of
the nebula has in general a much lower level of polarization, less than
30\% of the maximum. The polarized fraction for the case
$F(\psi)=\sin{(\psi)}$ is shown in Fig.~\ref{fig:faxi}.  The polarized
fraction reaches the theoretical maximum on axis, where, because of the
symmetry of the configuration, there are no depolarization effects. It
then drops to  zero as one approaches the region where the
relativistic polarization  angle swing appears, and then rises again
toward the CD, where however the luminosity vanishes. The other
cases have a similar pattern.

\begin{figure}
	\centering
	\includegraphics[bb=30 90 550 330,width=.45\textwidth,clip]{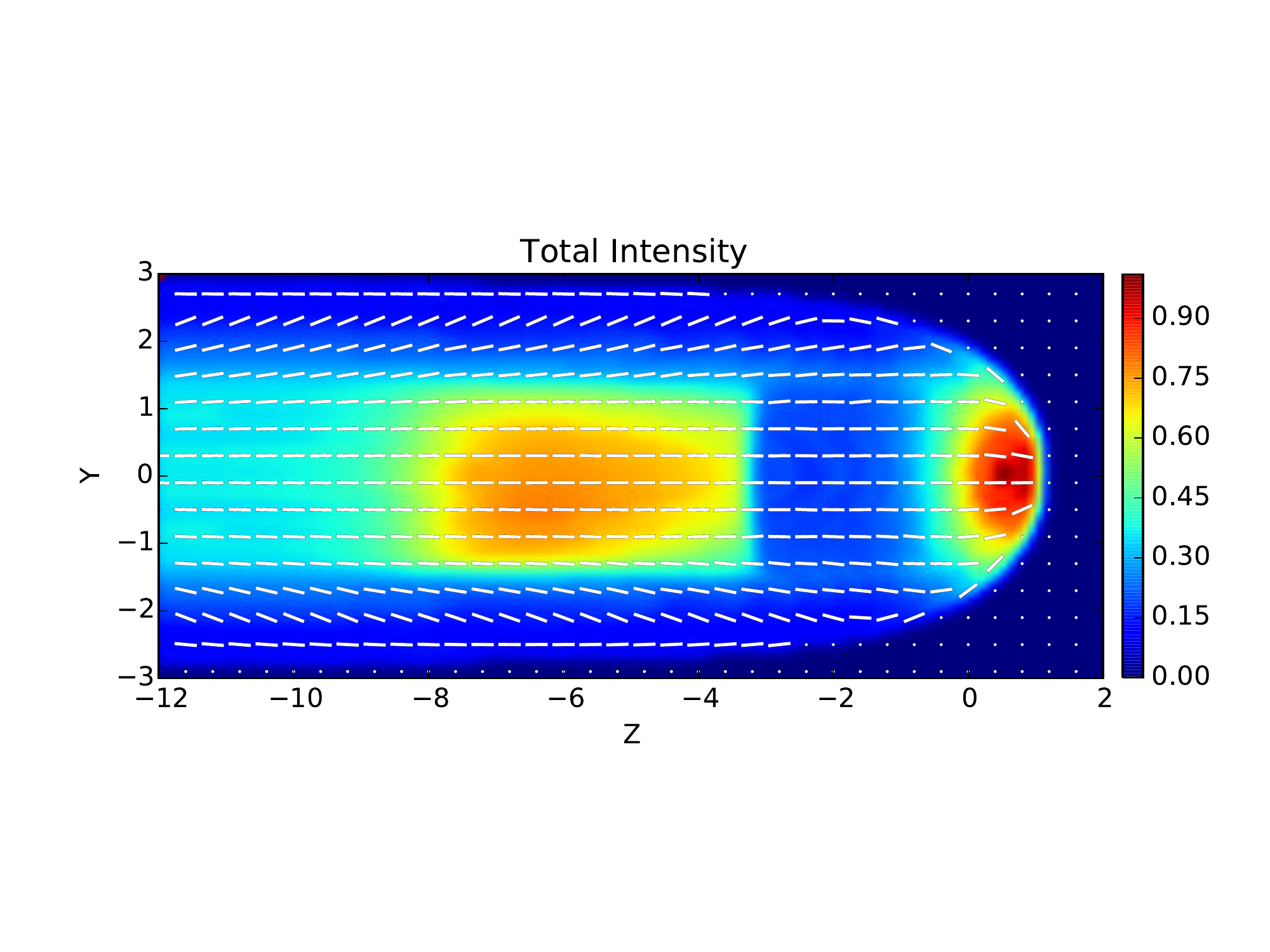}\\
	\includegraphics[bb=30 90 550 330,width=.45\textwidth,clip]{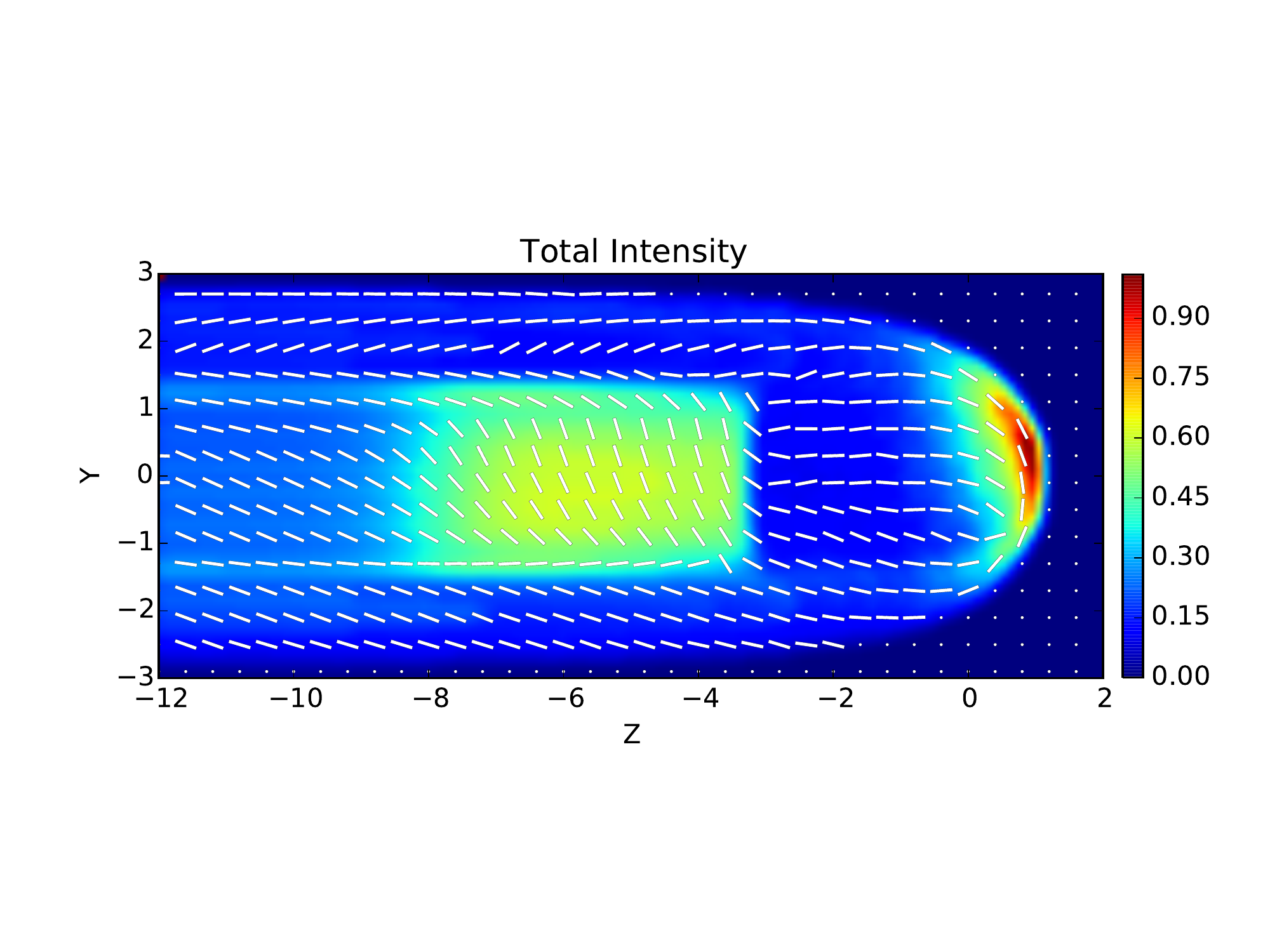}\\
        \includegraphics[bb=30 90 550 330,width=.45\textwidth,clip]{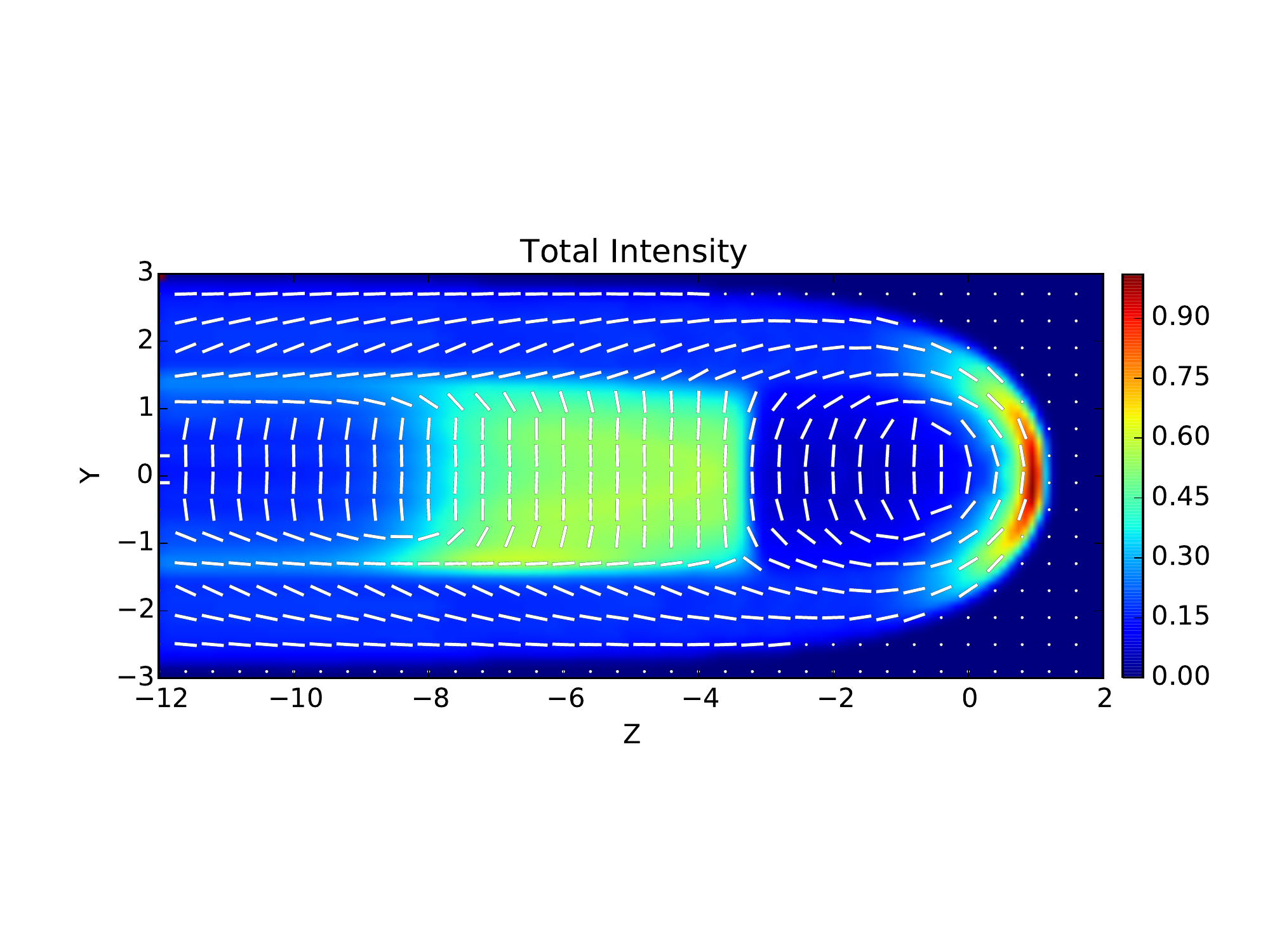}\\
	\caption{Total synchrotron intensity in the completely
          orthogonal case $\thi=90^\circ$ with $F(\psi)=\sin{(\psi)}$,
          but for different choices of the
          observer viewing angle: upper panel
          $\chi=0^\circ$, middle panel $\chi=45^\circ$, lower panel
          $\chi=90^\circ$. Intensity is
          normalized to the maximum. Dashes indicate the orientation of
          the polarization vector (not its amplitude). The pulsar is
          located in $Y=0$ $Z=0$.
     }
	\label{fig:int90}
\end{figure}

\begin{figure}
	\centering
	\includegraphics[bb=30 90 550 330,width=.45\textwidth,clip]{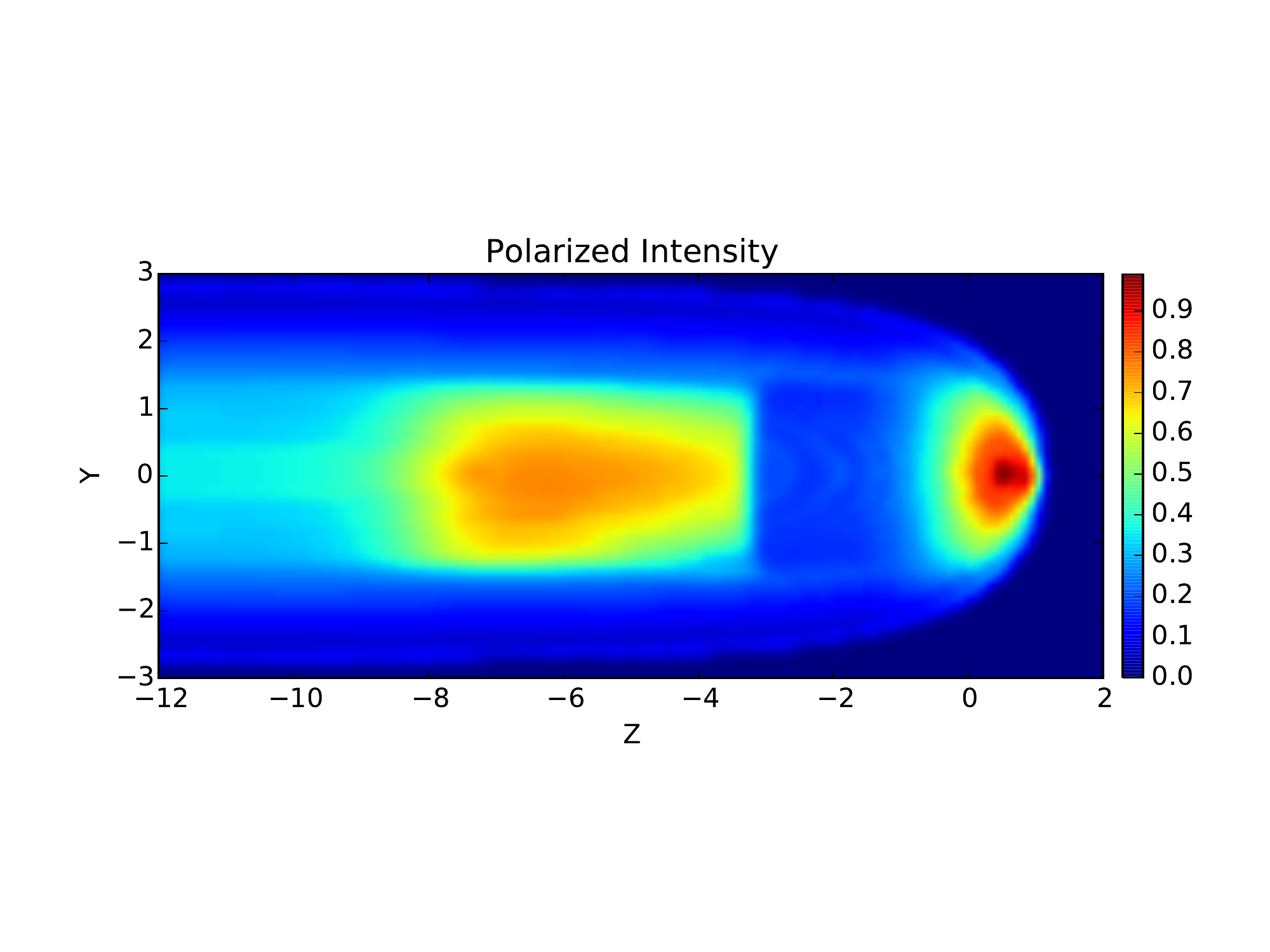}\\
	\includegraphics[bb=30 90 550 330,width=.45\textwidth,clip]{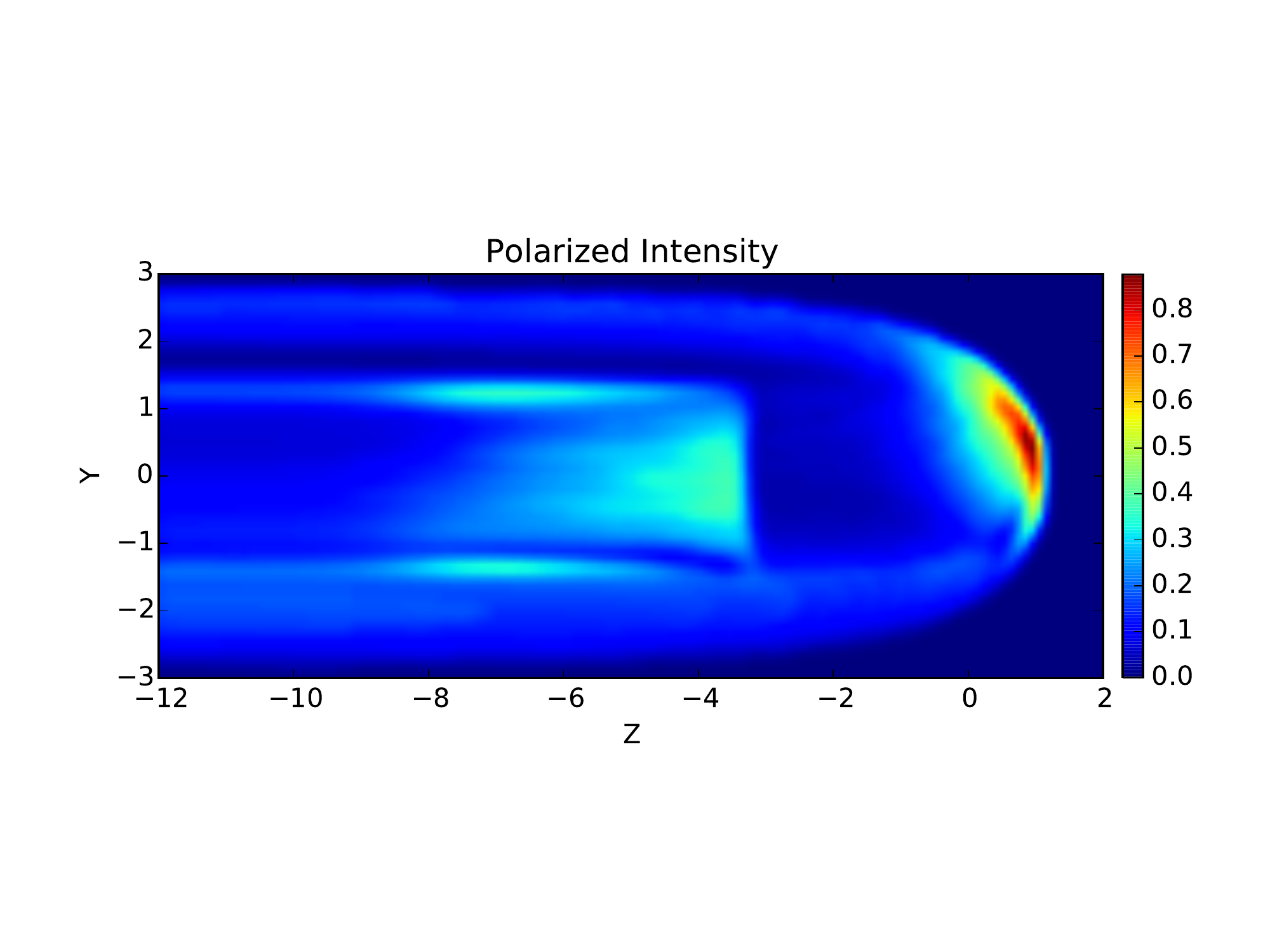}\\
        \includegraphics[bb=30 90 550 330,width=.45\textwidth,clip]{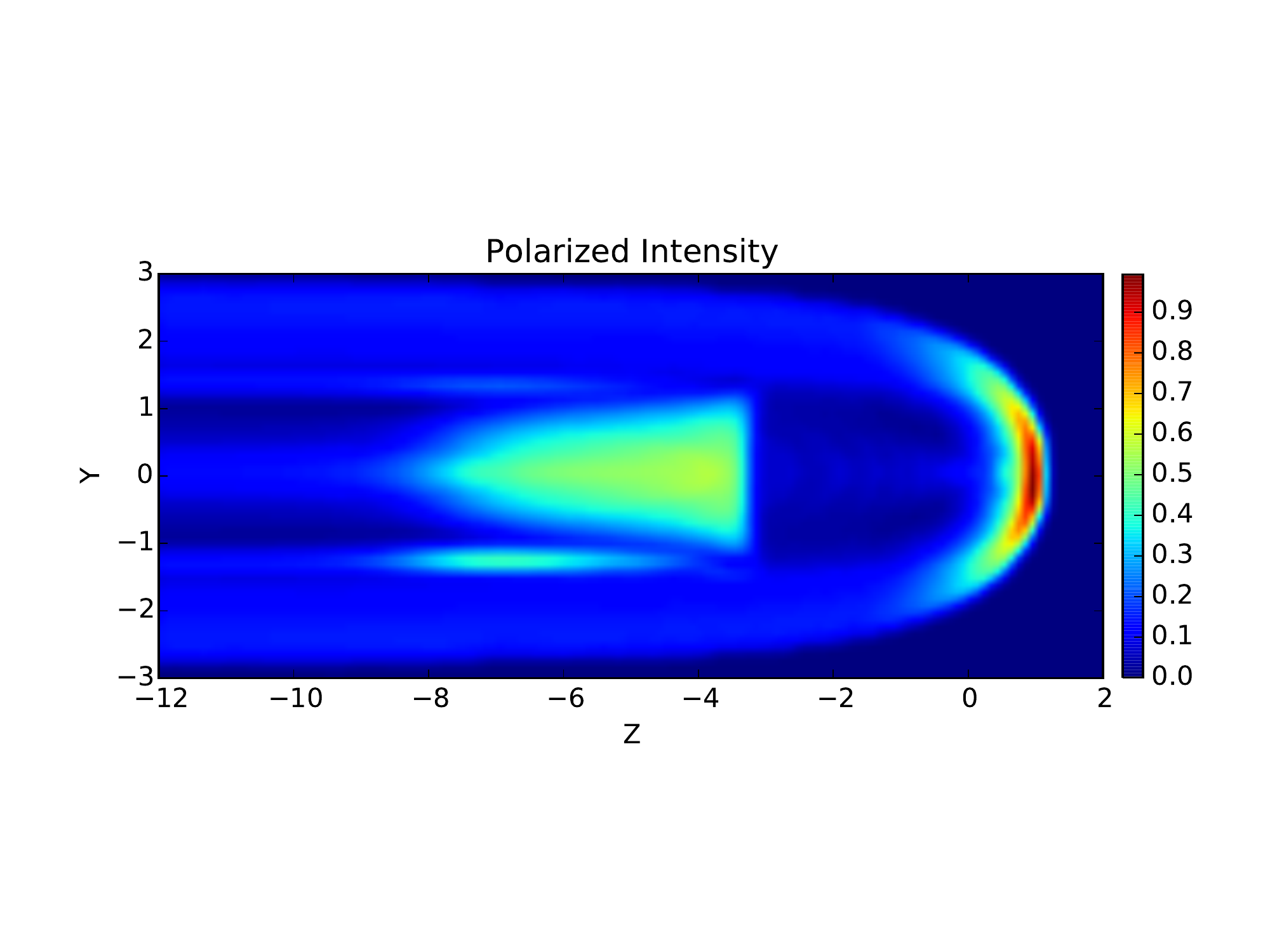}\\
	\caption{Same as Fig.~\ref{fig:int90} but for polarized
          intensity. Maps are normalized to the maximum of the total intensity.
     }
	\label{fig:pol90}
\end{figure}

\begin{figure}
	\centering
	\includegraphics[bb=30 90 550 330,width=.45\textwidth,clip]{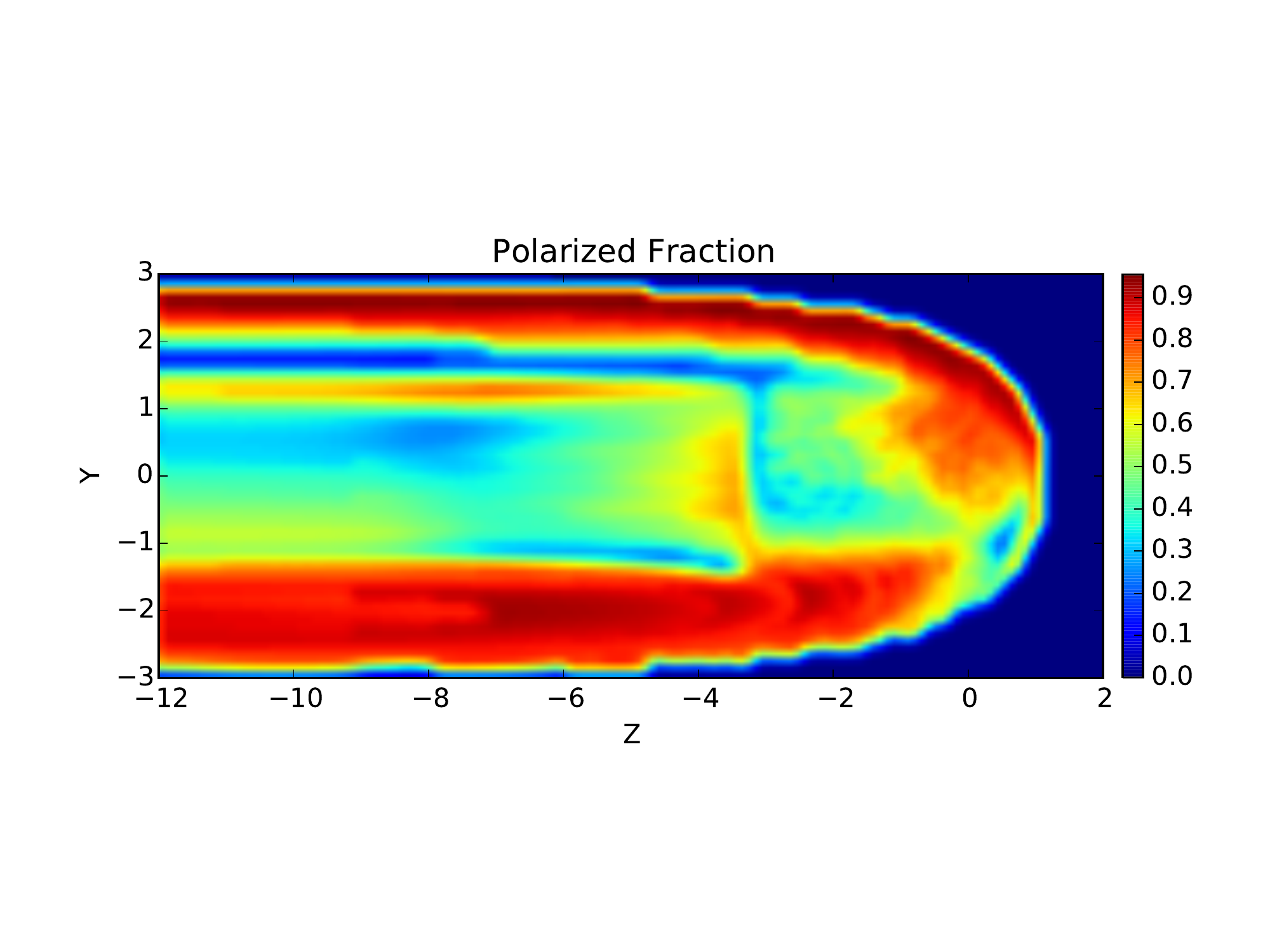}\\
	\includegraphics[bb=30 90 550 330,width=.45\textwidth,clip]{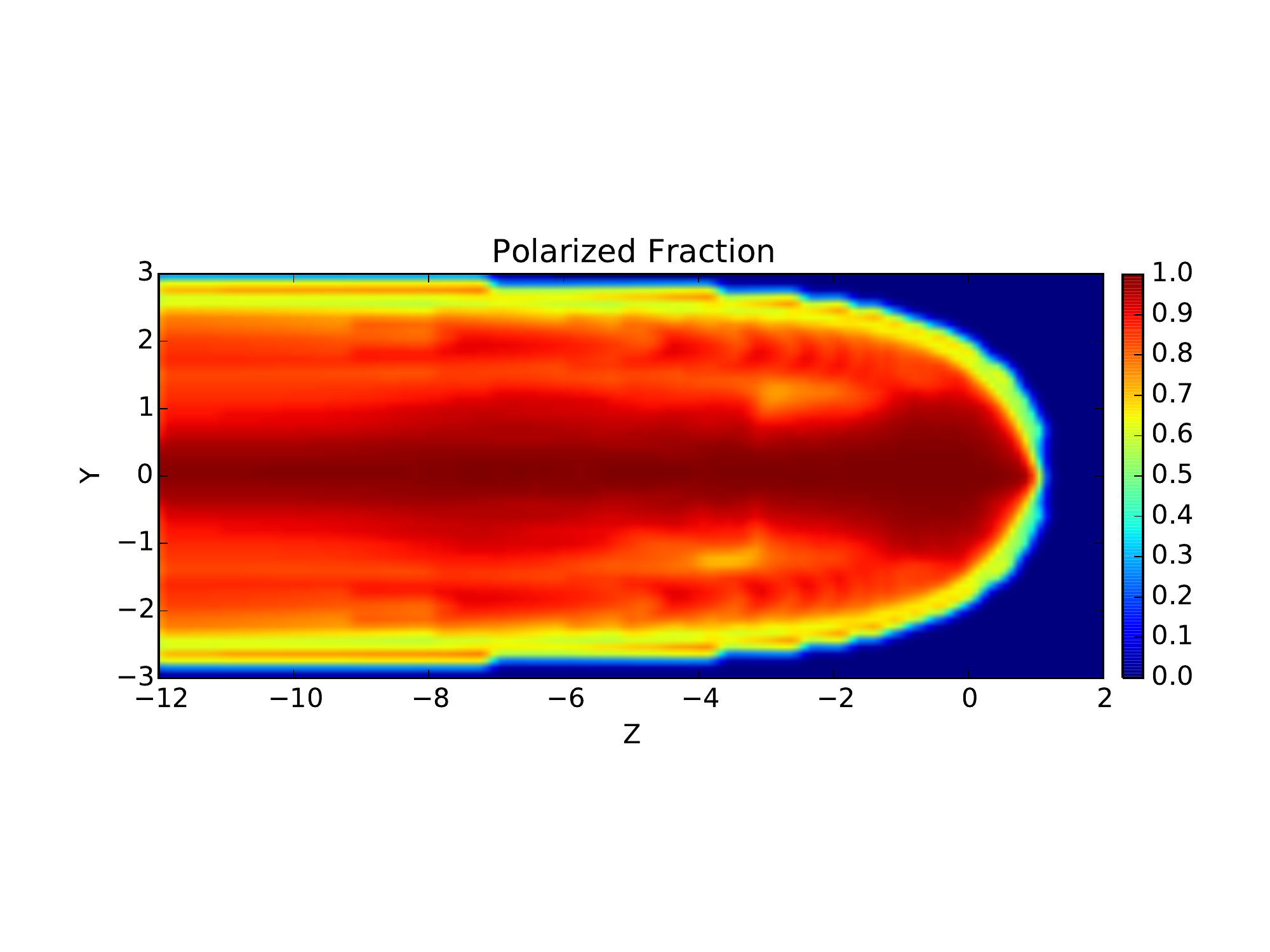}\\
	\caption{Polarized fraction for the case $\thi=0$,
          $F(\psi)=\sin{(\psi)}$, and two different values of the
          viewing angle: upper panel $\chi=45^\circ$, lower panel $\chi=0^\circ$.
     }
	\label{fig:f45}
\end{figure}

In Fig.s~\ref{fig:int90} and \ref{fig:pol90} we show the total surface
brightness, polarization angle and polarized intensity, for the fully
orthogonal case $\thi=90^\circ$. We limit our model to the canonical $F(\psi)=
\sin{(\psi)}$ profile for the magnetic field distribution in the wind,
and investigate the appearance of the nebula at three different
viewing angles $\chi=0,45$ and $90$ degrees. For $\thi=90^\circ$ the system
is symmetric with respect to the direction transverse to the $Z$-axis
both for $\chi =0^\circ$ and $\chi=90^\circ$. With respect to the axisymmetric
case $\thi=0^\circ$ we note immediately that the region corresponding to the
inner slow channel downstream of the Mach Disk, is brighter, while the
outer parts of the nebula are far weaker. The brightness profile in the
head is now strongly dependent of the viewing orientation, becoming
very arcuate for $\chi=90^\circ$, while at intermediate viewing angles
it shows a clear asymmetry, as expected. In practice we are observing
how the emission from the magnetic chimney changes, depending on
orientation.  Even the orientation of the polarization angle is as
expected. Interestingly for $\chi=90^\circ$ the observed polarization
angle follows the shape of the CD in the head. In this case, the
inclination of the polarization in the tail maps the relative angle
between the observer and the pulsar spin axis. On the other hand major
differences are evident in the map of polarized intensity. For
$\chi=0^\circ$ the polarized intensity in the region corresponding to the slow
inner flow channel can reach values up to 70\% of the total brightens
maximum, significantly larger than in the axisymmetric $\thi=0^\circ$ case,
and even in the asymptotic tail polarizations of 40\% of the maximum
are possible. In this case the pattern of polarized fraction is similar
(because of symmetry) to the one shown in Fig.~\ref{fig:faxi}. For
other viewing angles instead, the polarized intensity can be
quite smaller. For $\chi=45^\circ$ the inner slow channel appears
weak in polarized light. For $\chi=90^\circ$ the polarized intensity in
the inner flow channel is still high (comparable to the axysymmetric
case). Note that the
regions with high polarized intensity located in the tail at $Y=\pm 1.3$ are due to
the local strong shear between the inner slow channel and the outer
fast flow, that leads to a strong distortion of the local magnetic
field. In terms of polarized fraction the results for $\chi=45^\circ$ and
$\chi=90^\circ$ are shown in Fig.~\ref{fig:f45}. The tail tends to become
   progressively more polarized, and at $\chi=90^\circ$ results
   uniformly in excess of 70\% of the theoretical maximum.

\begin{figure}
	\centering
	\includegraphics[bb=30 90 550 330,width=.45\textwidth,clip]{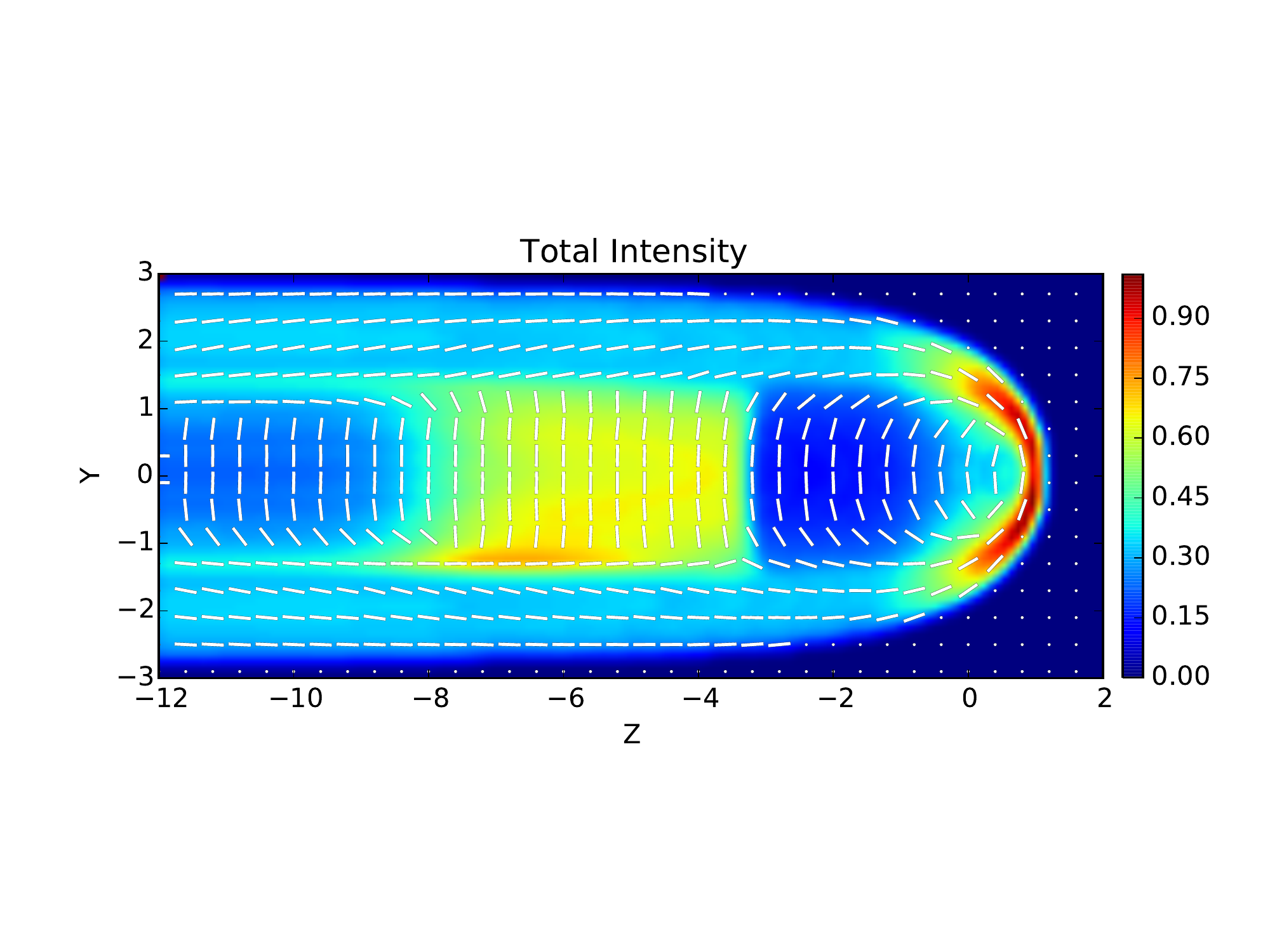}\\
	\includegraphics[bb=30 90 550 330,width=.45\textwidth,clip]{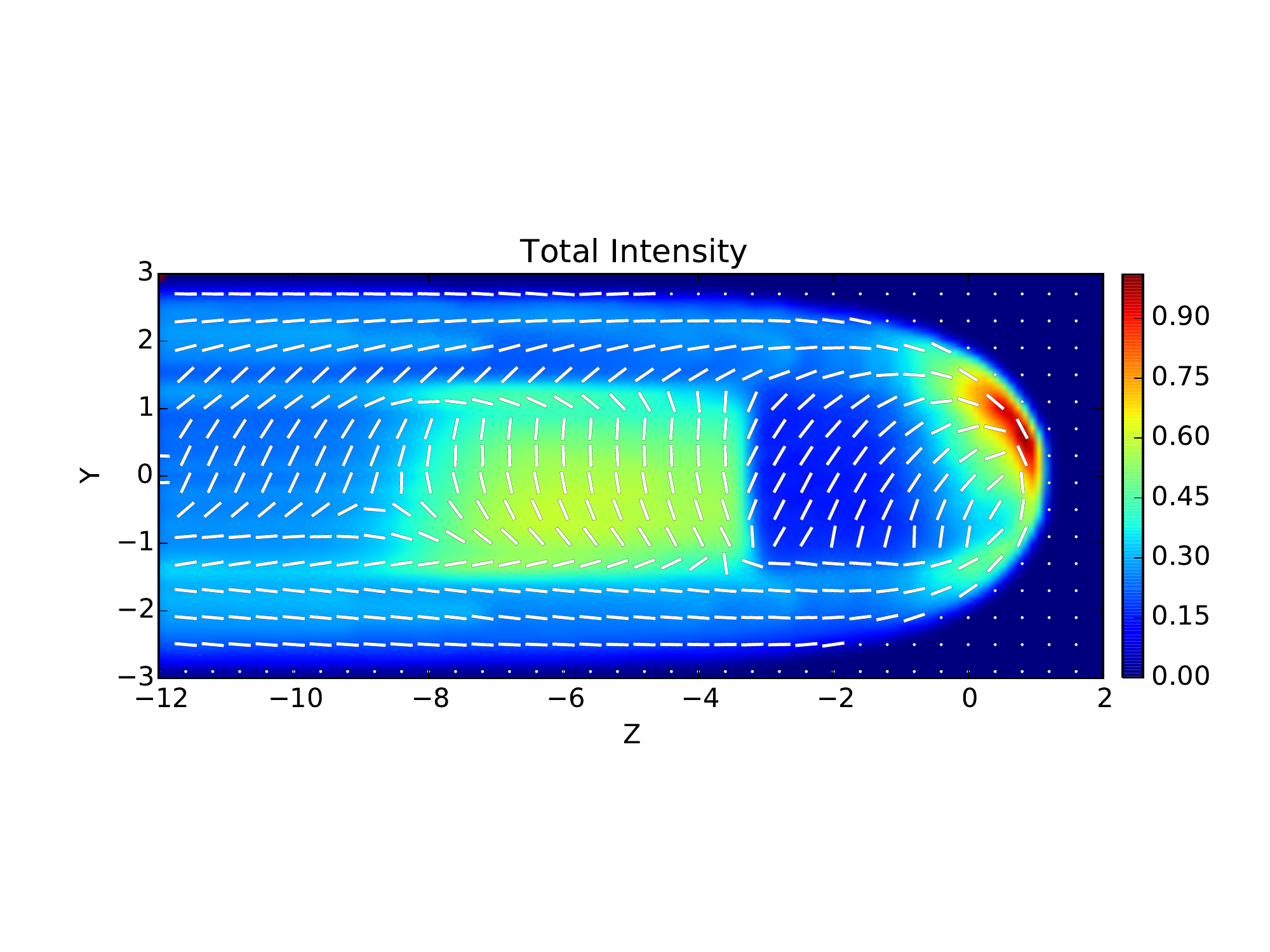}\\
        \includegraphics[bb=30 90 550 330,width=.45\textwidth,clip]{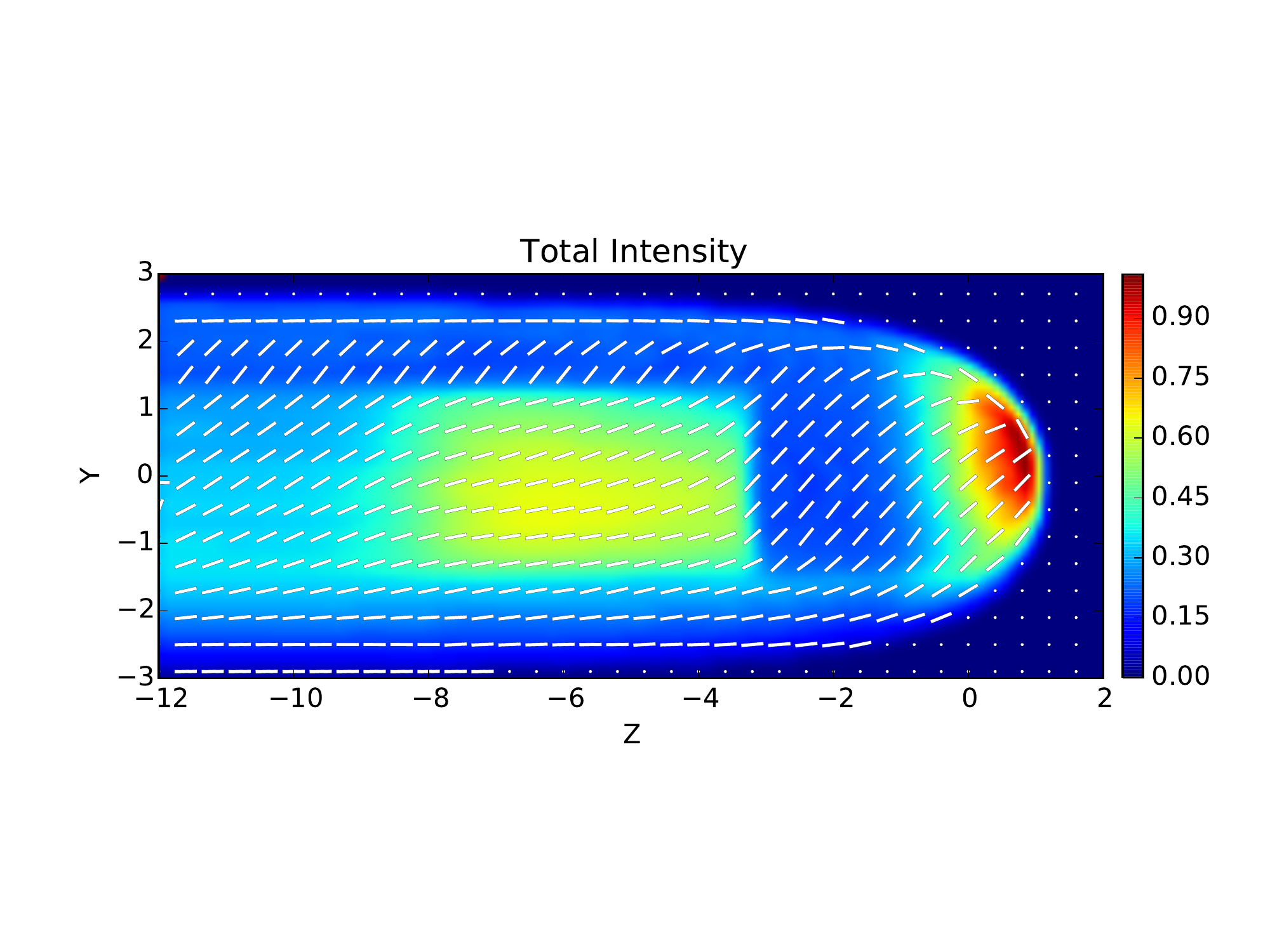}\\
        \includegraphics[bb=30 90 550 330,width=.45\textwidth,clip]{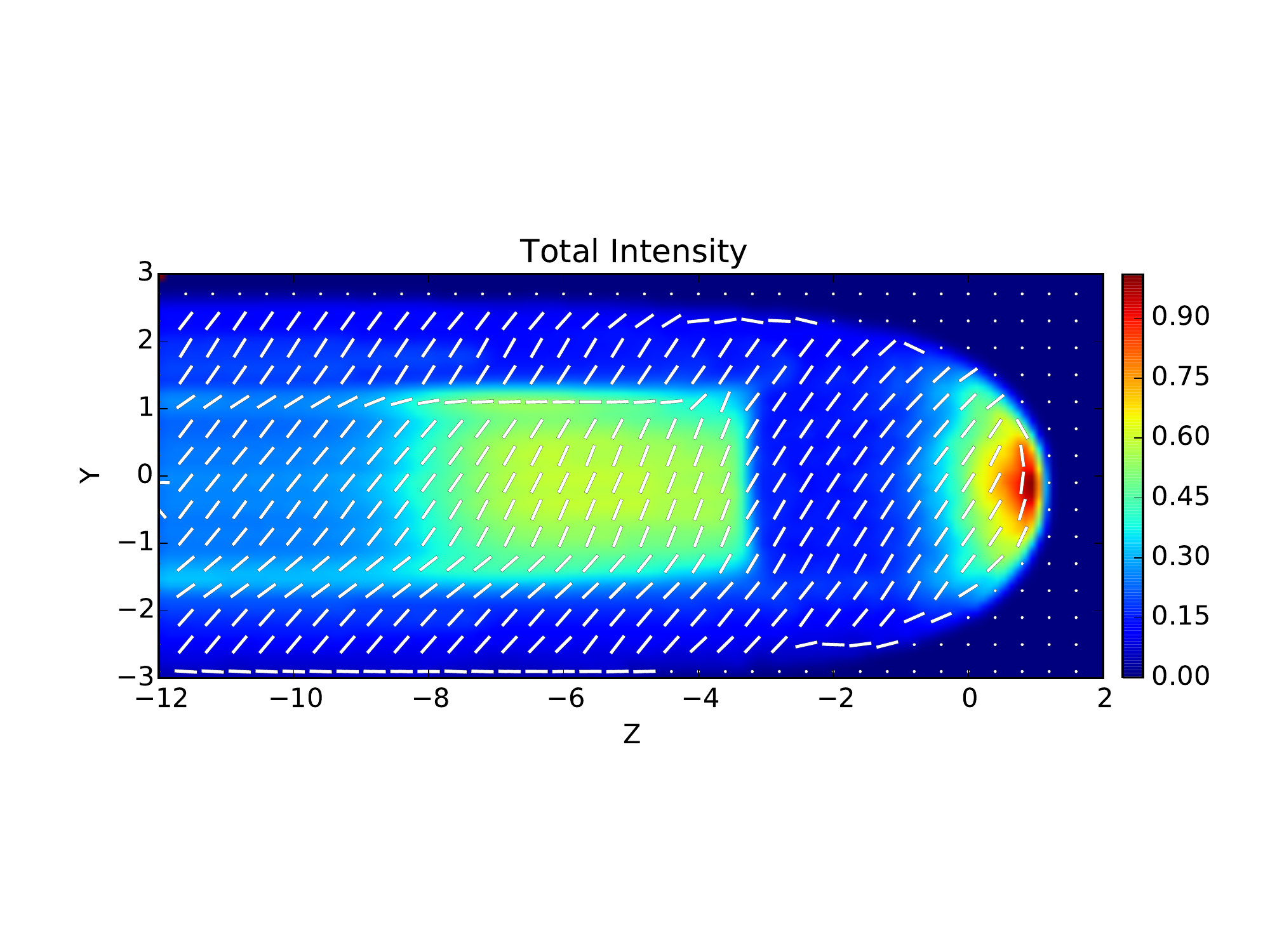}\\
        \includegraphics[bb=30 90 550 330,width=.45\textwidth,clip]{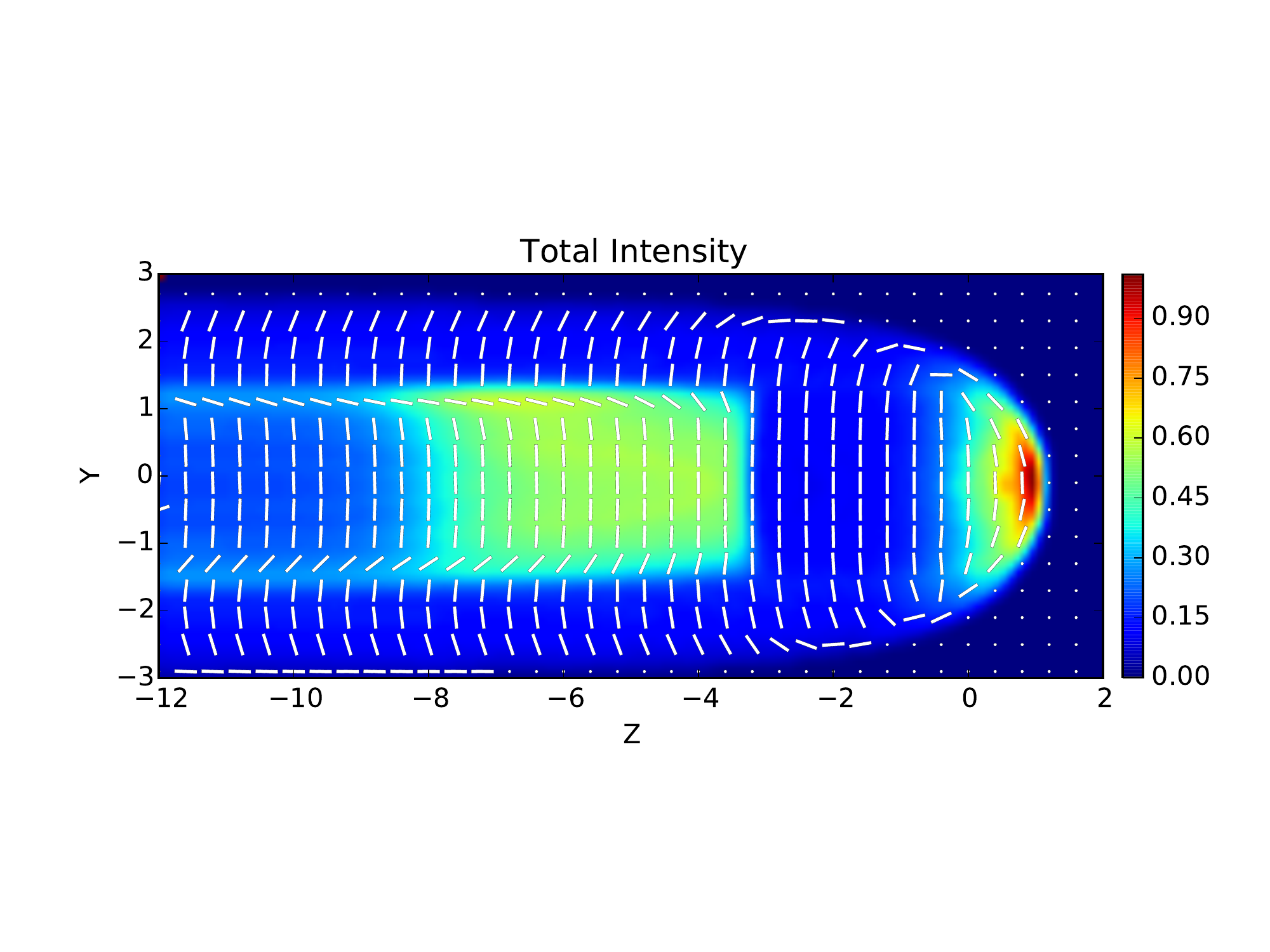}\\
	\caption{Total synchrotron intensity in the inclined case $\thi=45^\circ$ with $F(\psi)=\sin{(\psi)}$,
          but for different choices of the
          observer viewing angle: from the upper panel
       to the  lower one
          $\chi=90,45,0,-45,-90$ degrees. Intensity is
          normalized to the maximum. Dashes indicate the orientation of
          the polarization vector (not its amplitude). The pulsar is
          located in $Y=0$ $Z=0$.
     }
	\label{fig:int45}
\end{figure}

\begin{figure}
	\centering
	\includegraphics[bb=30 90 550 330,width=.45\textwidth,clip]{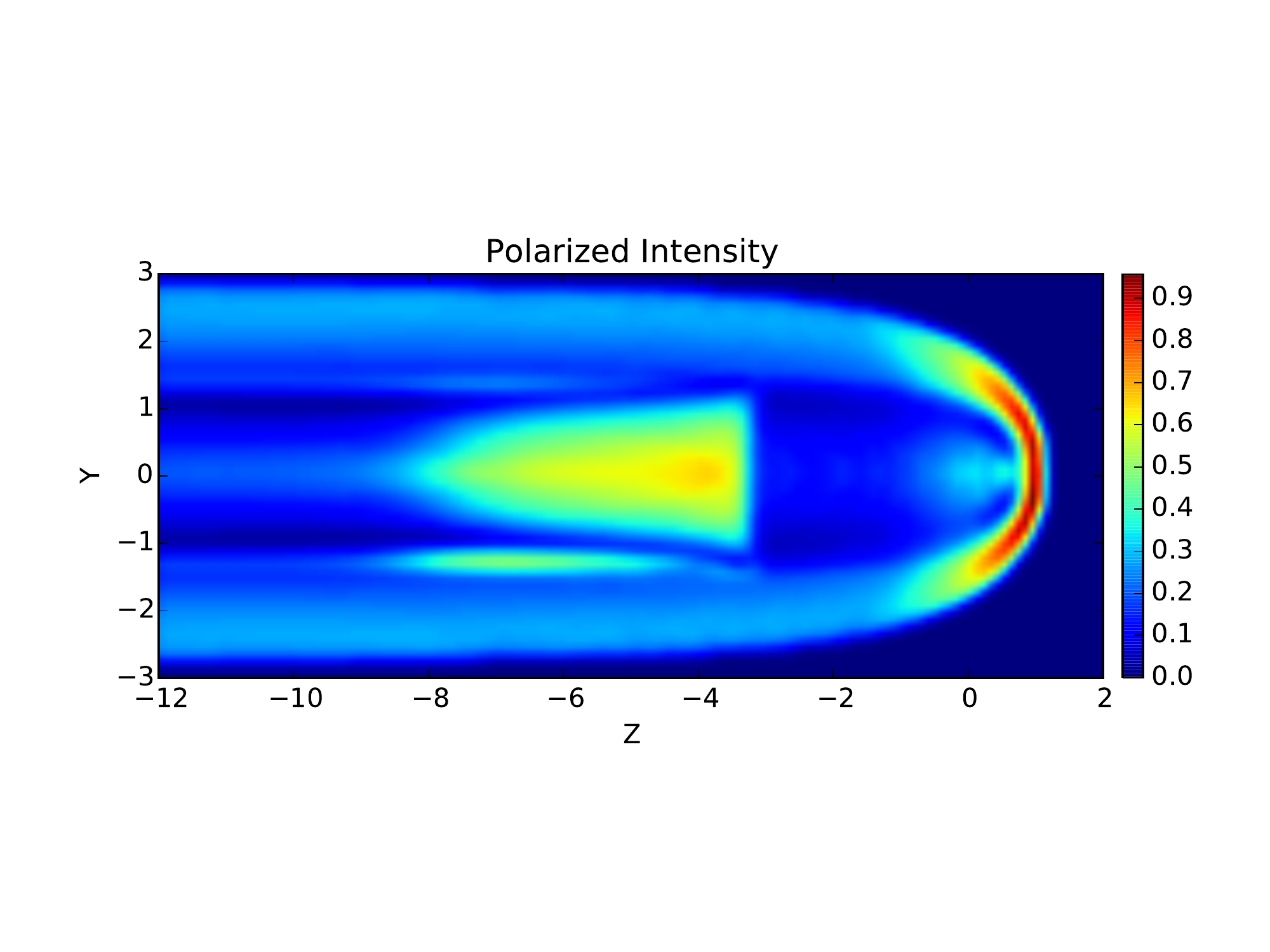}\\
	\includegraphics[bb=30 90 550 330,width=.45\textwidth,clip]{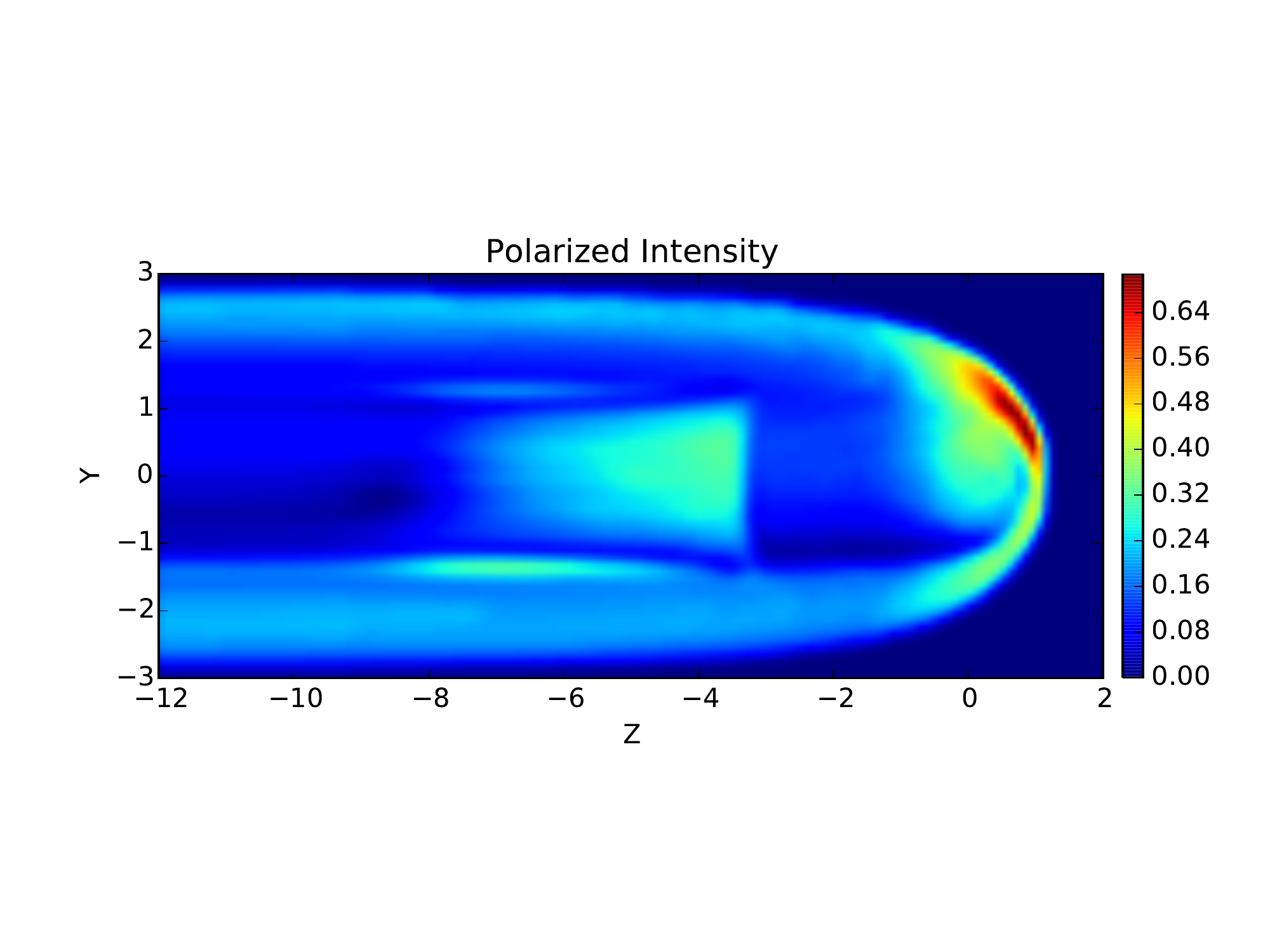}\\
        \includegraphics[bb=30 90 550 330,width=.45\textwidth,clip]{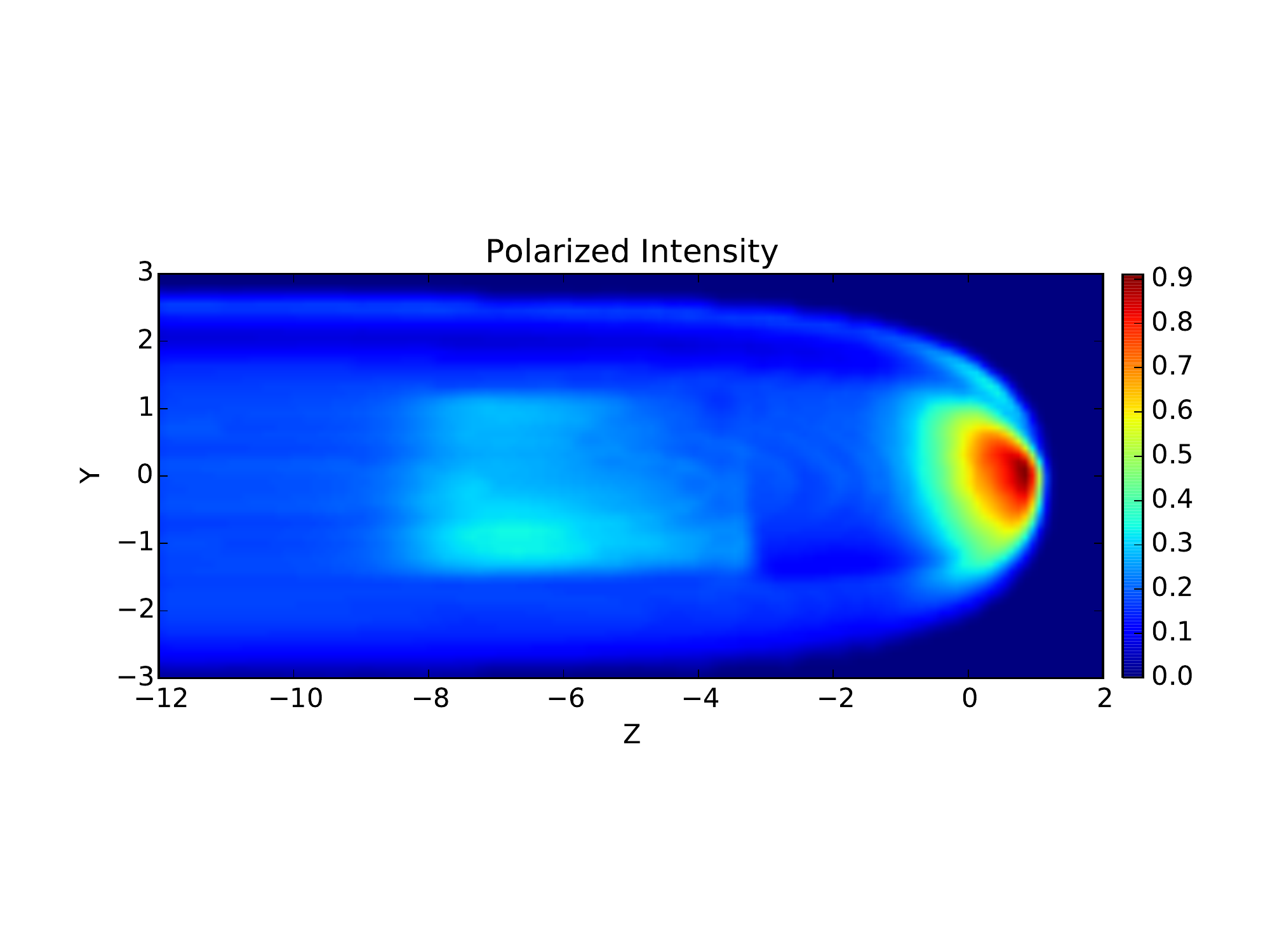}\\
        \includegraphics[bb=30 90 550 330,width=.45\textwidth,clip]{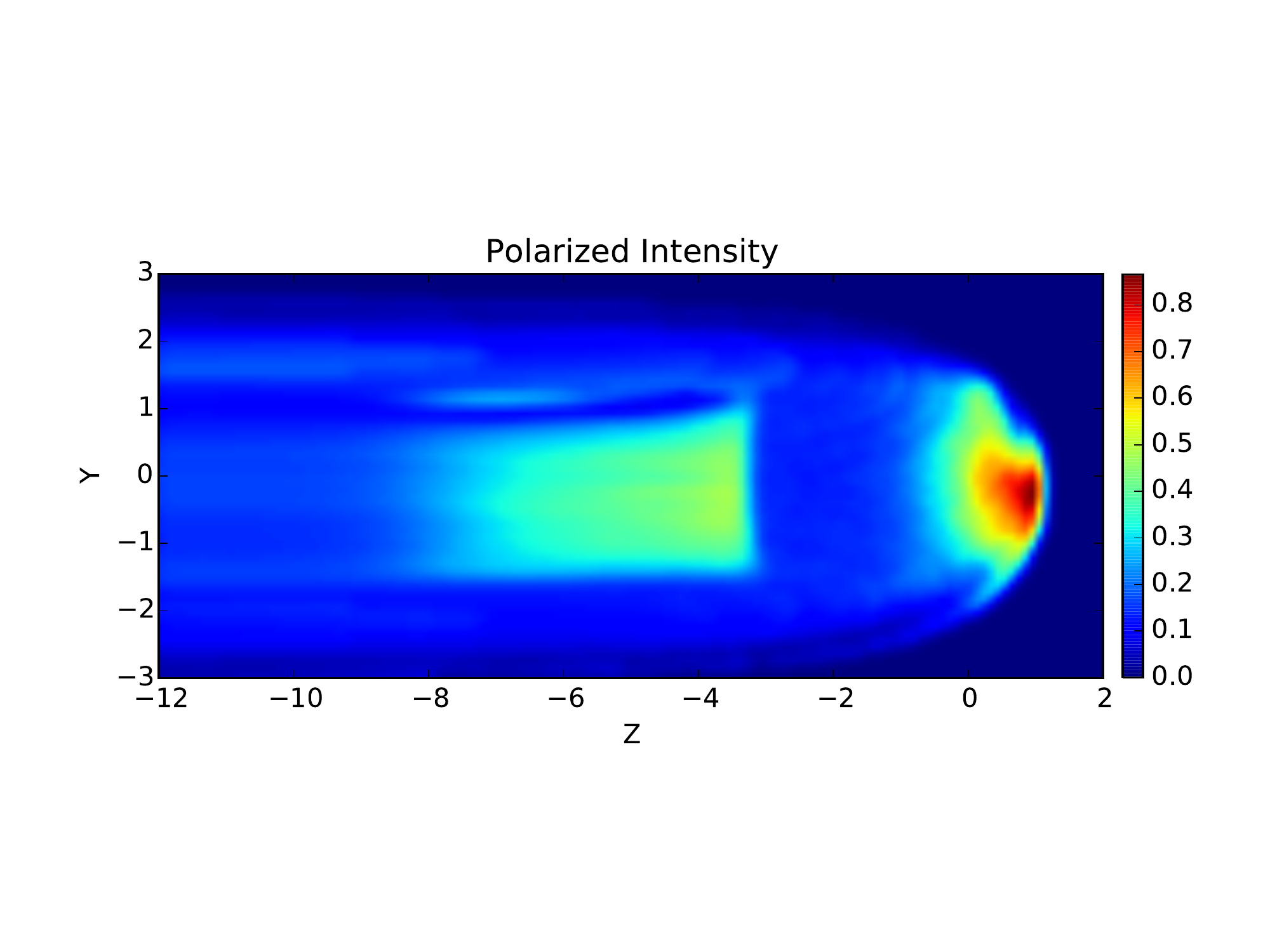}\\
        \includegraphics[bb=30 90 550 330,width=.45\textwidth,clip]{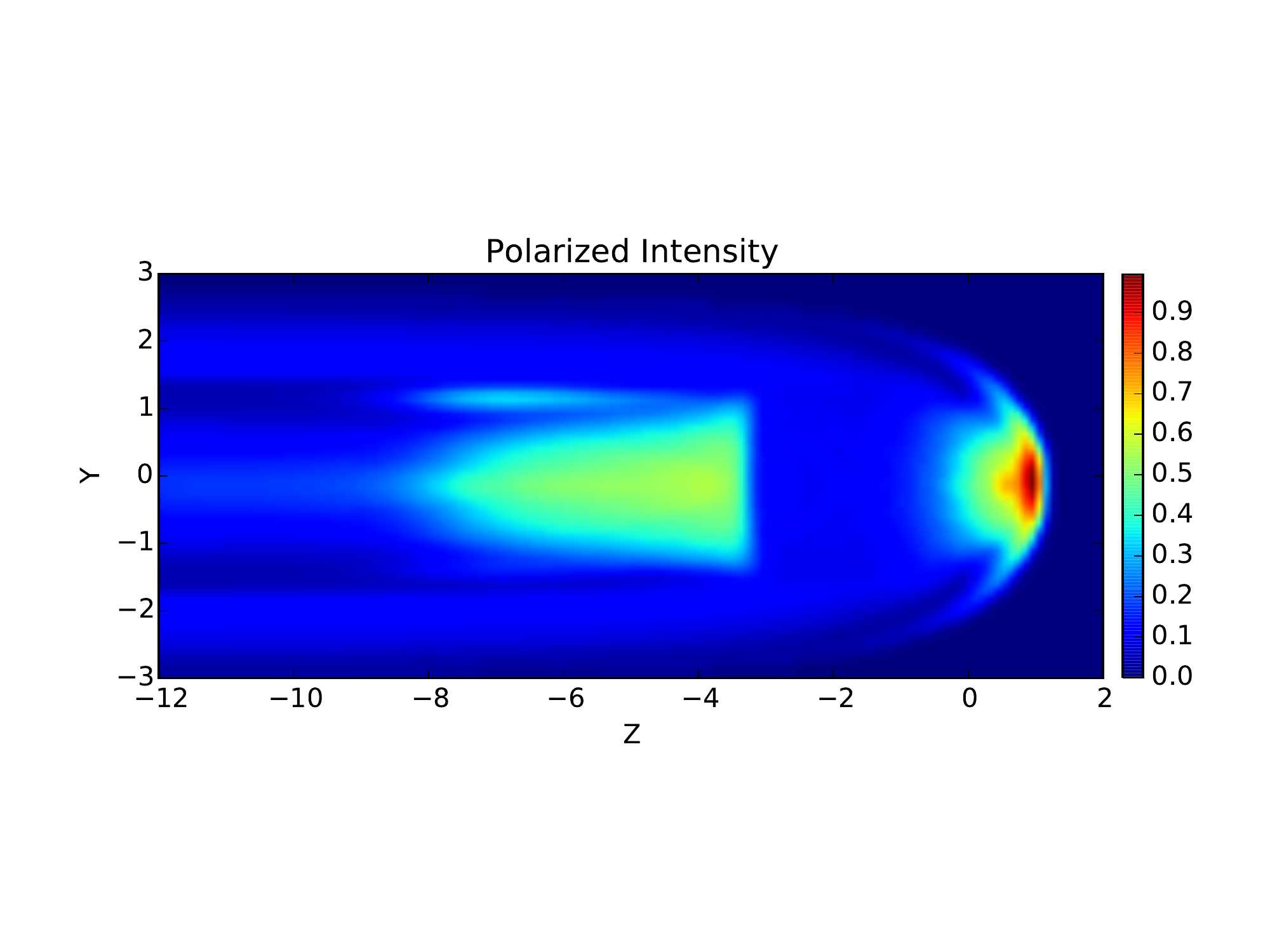}\\
	\caption{Same as Fig.~\ref{fig:int45} but for polarized
          intensity. Maps are normalized to the maximum of the total intensity.
     }
	\label{fig:pol45}
\end{figure}

In Fig.s~\ref{fig:int45} and \ref{fig:pol45} we show the total surface
brightness, polarization angle and polarized intensity, for the intermediate
 case $\thi=45^\circ$. We limit again our model to the canonical $F(\psi)=
\sin{(\psi)}$ profile for the magnetic field distribution in the wind,
and investigate the appearance of the nebula at five different
viewing angles $\chi=0,45,90,135,180$ degrees. In this case, in fact,
the presence of an inclined magnetic chimney defines a privileged
orientation breaking the up-down symmetry of the orthogonal case. It
is evident that the major differences due to the observer inclination
$\chi$ are in the brightness profile of the head that changes from
arch-like to more concentrated. It is also evident that the level of
asymmetry is maximal when the magnetic chimney points toward the
observer than when it points away. Also the polarization angle
experiences a similar change from orthogonal to quasi-parallel to the
$Z$-axis. Looking at the polarized intensity we see that, in the tail
the region corresponding to the inner slow flow channel in the cases
$\chi=0^\circ$ and $\chi=180^\circ$, it shows a high level of polarized intensity $\sim
60$\% of the total brightness maximum, less than in the orthogonal case but
higher that for the completely axisymmetric case. However, for other
 inclinations the polarized intensity drops substantially  and the
tails looks almost uniform in polarized emission  for
$\chi=90^\circ$. In terms of polarized fraction we find that for $\chi
=0^\circ$ and $\chi=180^\circ$ the polarized fraction on axis reaches its theoretical
maximum (as expected given that at this observer inclination there are no
depolarization effects on axis). It also reaches values close to
70-80\% of the theoretical maximum close to the CD. For $\chi=90^\circ$
instead the polarized fraction in the tail appears uniform at about
50-60\% of the theoretical maximum.

\subsection{The role of the spectral index}

\begin{figure}
	\centering
	\includegraphics[bb=30 90 550 330,width=.45\textwidth,clip]{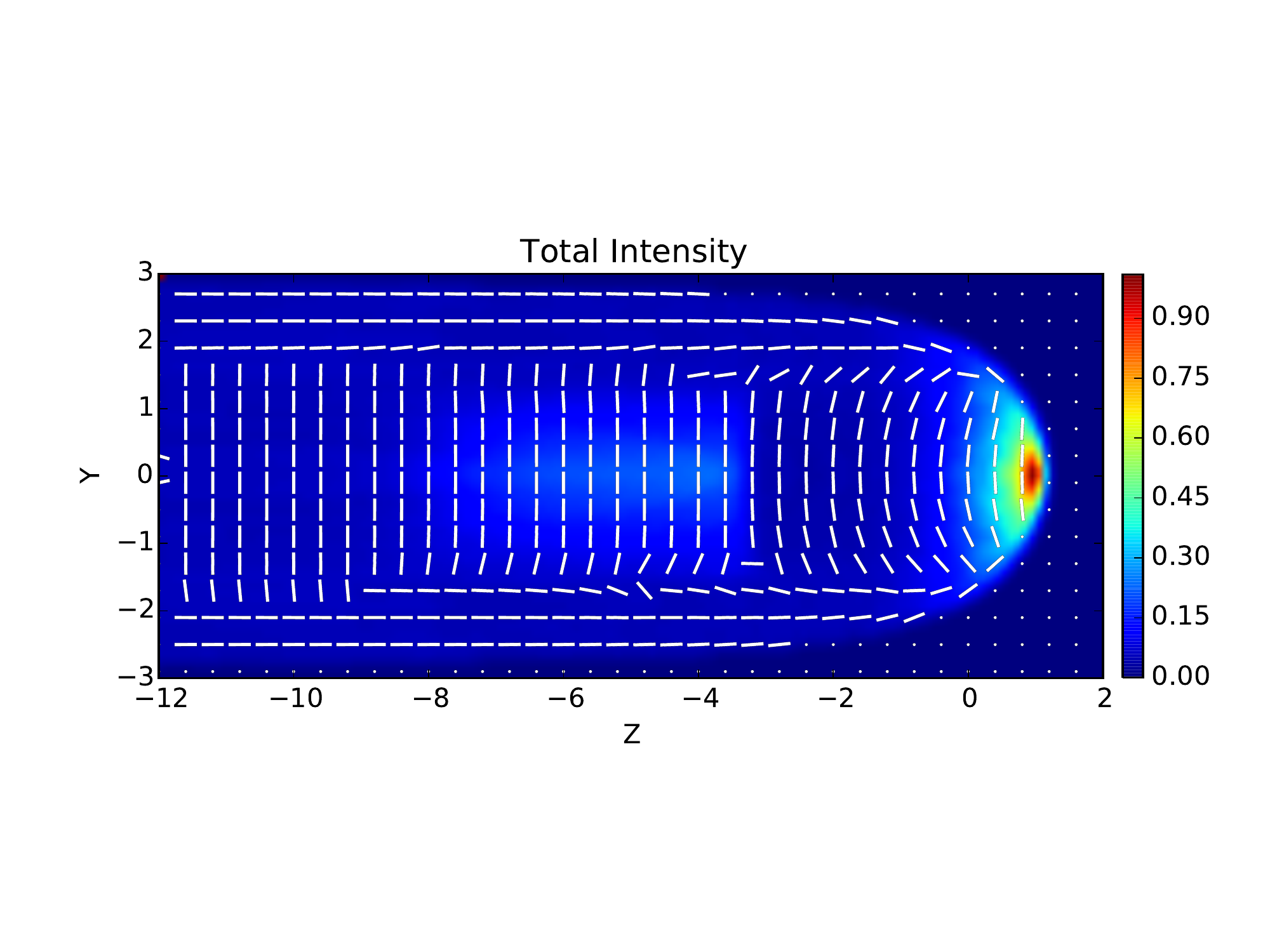}\\
	\includegraphics[bb=30 90 550 330,width=.45\textwidth,clip]{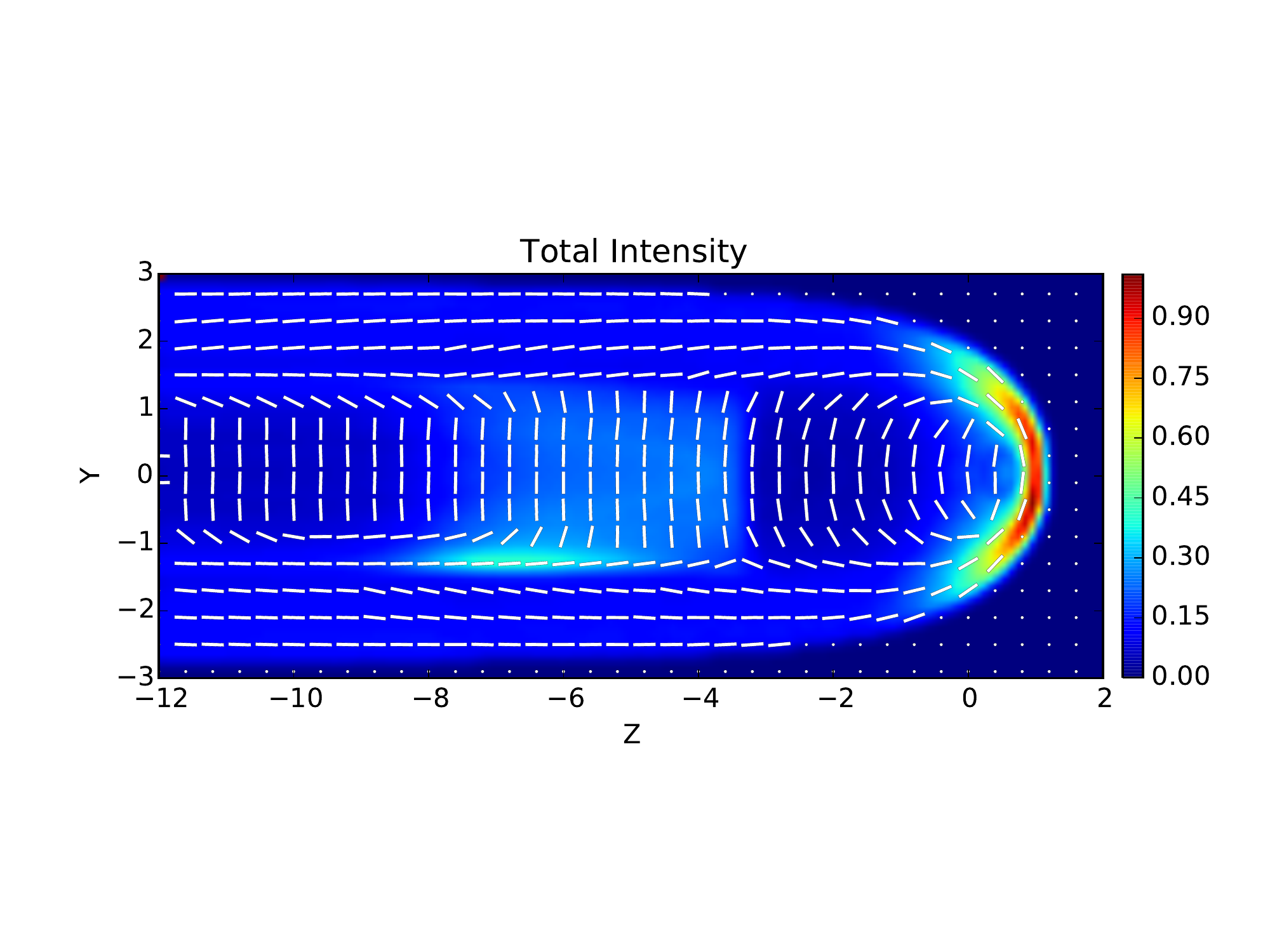}\\
        \includegraphics[bb=30 90 550 330,width=.45\textwidth,clip]{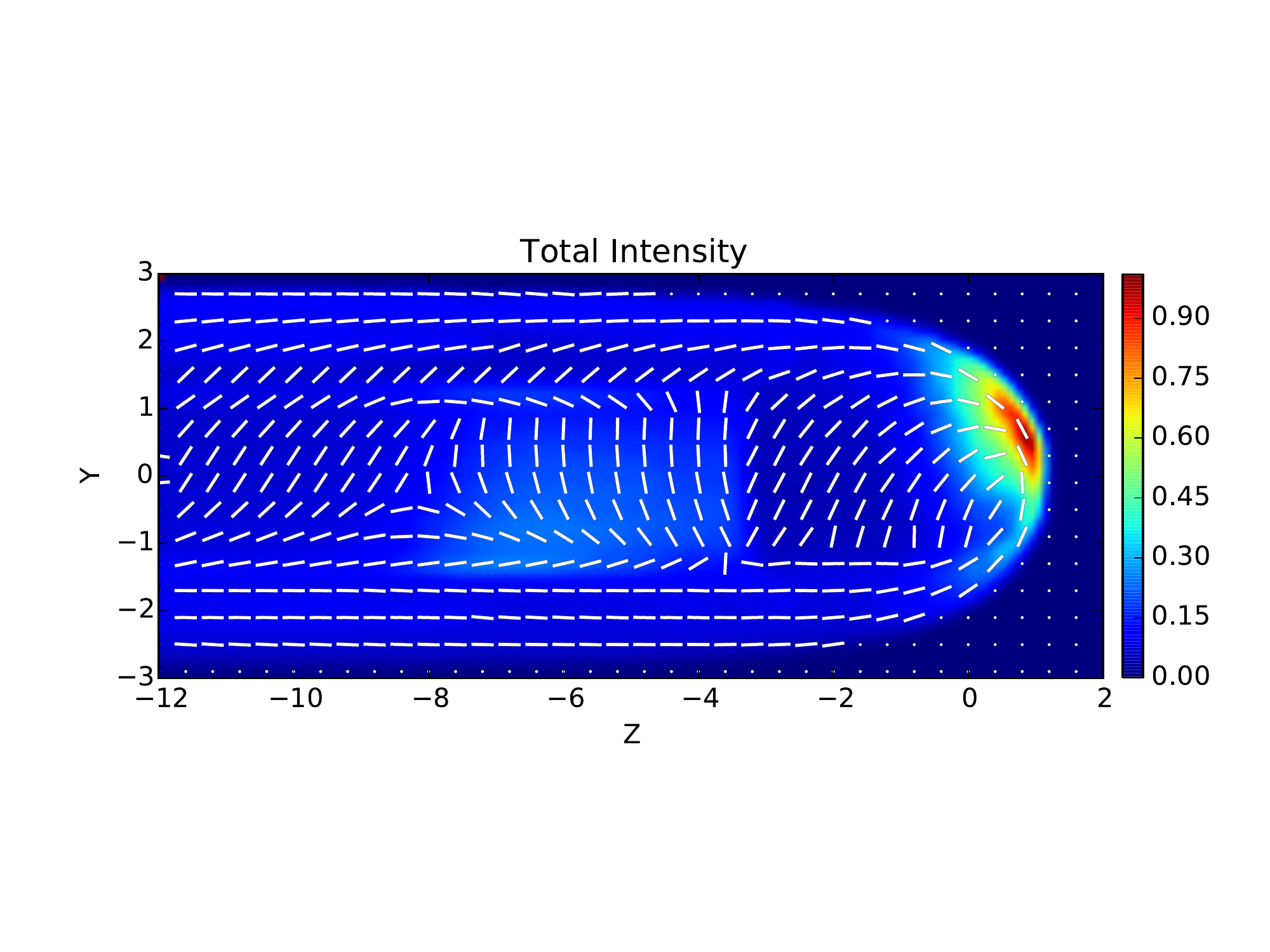}\\
	\caption{ Total intensity maps for a spectral index
          $\alpha=0.6$. Upper panel: completely axisymmetric case
          $\thi=0$, with $F(\psi)=const$ (see middle panel of
          Fig.~\ref{fig:intaxi} for comparison). Middle panel:
          inclined case $\thi=45^\circ$ with $F(\psi)=\sin{(\psi)}$
          and  $\chi=90$ ( see top panel of
          Fig.~\ref{fig:int45} for comparison).  Bottom panel:
          inclined case $\thi=45^\circ$ with $F(\psi)=\sin{(\psi)}$
          and  $\chi=45$ ( see second panel of
          Fig.~\ref{fig:int45} for comparison). 
     }
     \label{fig:spectral}
\end{figure}

 We briefly illustrate here how the emission maps change, varying the
spectral index $\alpha$. Observations suggest that the radio spectral
index should be quite flat. In Fig.~\ref{fig:spectral} we show the
total intensity map corresponding to the case $\alpha =0.6$, for a few
selected geometries. It is immediately evindent that, a steeper spectral
index, produces maps of total intensity (the same hold for the
polarized intensity), where the contrast between brighter and fainter
regions is enhanced, and the presence of a brighter region in the tail,
corresponding to the slow inner channel, is much less evident. The reason for this behaviour can be understood
by looking at Eq.~\ref{eq:spectral}, where is it evident that a
higher value of $\alpha$ will increase the emissivity toward the
observer from the regions of stronger field (the head) and the regions
of higher Doppler boosting (again in the head where the flow points
toward the observer), with respect to the tail. The first panel of
Fig.~\ref{fig:spectral}, shows a map in a configuration analogous to
the low magnetization case of \citet{Bucciantini_Amato+05a}, which
also shows a stronger emission in the head, with no evidence for a
brighter region in the tail corresponding to the slow moving
channel. However a direct comparison is not possible, because
\citet{Bucciantini_Amato+05a}, being mostly interested in the X-ray
properties,  adopt a pressure normalization in the
emissivity, and include the effect of synchrotron cooling. These two
extra terms enhance even more the head to tail brightness
ratio. Interestingly the choice of spectral index does not affect the
polarization structure.

\section{Conclusions}
\label{sec:conc}

We have introduced here a method based on Lagrangian tracers to compute
the structure and geometry of the magnetic field for a given velocity
field in 3D, and we have applied it to compute the structure of the
magnetic field expected inside a Bow Shock Pulsar Wind Nebula. This
approach allowed us to compute the full magnetic structure of the nebula
on a single CPU in just a few hours as opposed to full multidimensional
numerical models that require hundreds of hours on large scale
multiprocessor machines. A
simple analytical steady-state laminar flow pattern has been adopted to
describe the properties of the flow of the material injected by the
pulsar, and responsible for the non-thermal emission that is
observer. Despite the simplicity of this assumption we have shown
that a rich morphology in terms of radio emission maps and polarized
properties can arise due to the combination of different inclinations
of the pulsar spin axis, different distributions of the magnetic field
in the pulsar wind, and different orientations of the observer. 
  We caution the reader however, that more complex flow structures can
  arise, if an energy flux anisotropy of the wind is assumed, as
  suggested by magnetospheric moldels of pulsars.

We have shown that major asymmetries due to the orientation of the
spin axis or the observer are preferentially confined in the very
head, where the emission pattern can change substantially, while the
region of the tail  shows  generally a quite similar appearance, with major
differences being related to the intensity of the slow flow channel
that is expected to arise downstream of the Mach Disk. This shows that
bright regions downstream of the termination shock are likely
associated to slowly moving material (bright spots in the tail of known
radio BSPWNe could then be associated to shocks that slow down the flow). Given that the laminar
assumption, that for simplicity we have introduced, might not be
correct in the presence of strong shear, at the CD, one can in
principle use the brightness ratio of the tail vs the head  as a proxy
for the flow speed. Nebulae that show no major brightness contrast
are likely to be characterized by slow flow speed in the tail, that
could be the consequence either of shear and turbulence developed at
the CD or of mixing with the denser and slower ISM material. 
  However, we have also show that the brightness contrast is strongly
  dependent on the spectal index: a steeper spectrum enhancing the head
  to tail brightness ratio. So one should take care in interpreting
  observational results. We have
also investigated the polarization properties of the BSPWNe, and
verified that with a good approximation the orientation of the
polarization in the tail can be used as a good proxy for the
orientation of the magnetic axis, with polarization pattern that
ranges from aligned to the pulsar speed to orthogonal. The polarized
fraction is found to be in general quite high, with depolarization
being limited only to certain specific configurations. This could also
be used to constrain the level of turbulence in observed systems, by
comparing the observed polarized fraction with the expectation of a
fully laminar model. For example the Frying Pan G315.78-0.23
\citep{Ng_Bucciantini+12a} has a very ordered magnetic field aligned
with the tail (as in the upper panel of Fig.~\ref{fig:int90}), with a
polarized fraction of 40-50\%, and extending for several parsecs. Using the subgrid model by
\citet{Bandiera_Petruk16a} recently applied to PWNe
\citep{Bucciantini_Bandiera+17a,Bucciantini_Olmi18a}, we can estimate
a typical turbulence with $\delta B/B \sim 0.3$. On the other hand low
polarization systems like the Mouse G359.23-0.82
\citep{Yusef-Zadeh_Gaensler05a}, with polarized fraction below 10\%
are likely characterized by strong turbulence, in part already
evident in the rapid change of polarization pattern. Intermediate
cases with polarization $\sim$30\% \citep{Ng_Gaensler+10a} on the other hand suggest
a turbulence with  $\delta B/B \sim 1$.

Our model does not include dissipation, however it is evident from
the shape of the magnetic surfaces, that the magnetic chimney can be
strongly compressed and distorted in the head of a BSPWN, leading to
the formation of a region where the current density is high. This
region might be prone to reconnection and dissipation, leading to a
local heating and acceleration of particles. We cannot exclude that
bright thin non-symmetric features might form close to the
CD. This could be the case of the hard tails observed for example in
Geminga \citep{Posselt_Pavlov+17a}.

In this paper we have limited our study to the emission properties of
the BSPWNe. However the method we have introduced can be used to model
also the structure of the magnetic field of the ISM. We plan in the
future to extend this model to include the ambient magnetic field, in
order to evaluate which configurations, in terms of relative
orientation of the pulsar spin axis and ISM magnetic field, could be
promising for reconnection of the internal and external magnetic
field. This process has been invoked \citep{Bandiera08a} in order to explain
  bright X-ray features observed in the Guitar and Lighthouse nebulae
  \citep{Hui_Becker07a,Pavan_Bordas+14a} that emerge almost orthogonally from the head of the bow
  shock, far beyond the supposed position of the CD. The general idea
  is that particles can diffuse from the BSPWN to the ISM via loci
  at the CD of preferential reconnection. 

\section*{Acknowledgements}

The authors acknowledged support from the PRIN-MIUR project prot. 2015L5EE2Y "Multi-scale simulations of high-energy astrophysical plasmas".

\bibliography{Bib}{}
\bibliographystyle{mn2e}

\appendix
\section{Flow field}
\label{sec:app}
We provide here the analytical expression for the direction of the
velocity field used in region B and C of the nebula, in a cylindrical
reference frame $(R,Z)$. with the $Z$-axis pointing along the pulsar
speed. The pulsar speed into the ISM has norm $V$. 
\begin{align}
\frac{v_Z}{v}&=\frac{\frac{Z}{\left(R^2+Z^2\right)^{3/2}}-\frac{V}{c}}{\sqrt{\left(\frac{Z}{\left(R^2+Z^2\right)^{3/2}}-\frac{V}{c}\right)^2+\frac{\left(R^{3/10}+1\right)^2 R^2}{
   \left(R^2+Z^2\right)^3}}}\\
\frac{v_R}{v}&=\frac{\left(R^{3/10}+1\right) R}{\left(R^2+Z^2\right)^{3/2} \sqrt{\left(\frac{Z}{ \left(R^2+Z^2\right)^{3/2}}-\frac{V}{c}\right)^2+\frac{\left(R^{3/10}+1\right)^2 R^2}{
   \left(R^2+Z^2\right)^3}}}
\end{align}

\end{document}